\documentclass[11pt,a4paper]{article}

\usepackage[T1]{fontenc}
\usepackage{lmodern}

\usepackage{amsmath}
\usepackage{calc}
\usepackage{amssymb}
\usepackage{hyperref}
\usepackage{graphicx}
\graphicspath{ {./figs/} }
\usepackage{bbold}
\usepackage{color}
\usepackage{geometry}
\usepackage{setspace}
\usepackage{xspace}
\usepackage{bbding}
\usepackage{mathtools}
\usepackage{array}
\usepackage{listings}
\usepackage{kvoptions}
\usepackage[dvipsnames]{xcolor}
\usepackage{textcomp}
\usepackage{cleveref}
\usepackage{booktabs}

\newcommand{\de}[0]{\text{d}}

\newcommand{\elena}[0]{\textsf{ELENA}\xspace}
\newcommand{\python}[0]{\textsf{Python}\xspace}

\newcommand{\CT}[0]{\textsf{CosmoTransitions}\xspace}

\usepackage{listings}
\usepackage{tikz}

\definecolor{solarized@base03}{HTML}{002B36}
\definecolor{solarized@base02}{HTML}{073642}
\definecolor{solarized@base01}{HTML}{586e75}
\definecolor{solarized@base00}{HTML}{657b83}
\definecolor{solarized@base0}{HTML}{839496}
\definecolor{solarized@base1}{HTML}{93a1a1}
\definecolor{solarized@base2}{HTML}{EEE8D5}
\definecolor{solarized@base3}{HTML}{FDF6E3}
\definecolor{solarized@yellow}{HTML}{B58900}
\definecolor{solarized@orange}{HTML}{CB4B16}
\definecolor{solarized@red}{HTML}{DC322F}
\definecolor{solarized@magenta}{HTML}{D33682}
\definecolor{solarized@violet}{HTML}{6C71C4}
\definecolor{solarized@blue}{HTML}{157AC1}
\definecolor{solarized@cyan}{HTML}{2AA198}
\definecolor{solarized@green}{HTML}{859900}
\definecolor{darkred}{HTML}{550003}
\definecolor{darkgreen}{HTML}{00AA00}
\definecolor{orchid}{HTML}{AF06F5}

\lstdefinestyle{python}
{
  language=Python,
  basicstyle=\footnotesize\ttfamily,
  basewidth={0.53em,0.44em},
  numbers=none,
  tabsize=2,
  breaklines=true,
  escapeinside={@}{@},
  showstringspaces=false,
  numberstyle=\tiny\color{solarized@base01},
  keywordstyle=\color{solarized@blue},
  stringstyle=\color{solarized@red}\ttfamily,
  identifierstyle=\color{solarized@blue},
  commentstyle=\color{purple},
  emphstyle=\color{green},
  frame=single,
  rulecolor=\color{solarized@base2},
  rulesepcolor=\color{solarized@base2},
  literate = {~}{\customtilde}1
             {\ as\ }{{\color{blue}\ as\ \color{black}}}3
             {.set}{{\color{black}.}{\color{darkred}set}}4
}
\lstnewenvironment{lstpy}{\lstset{style=python}}{\lstset{style=python}}
\newcommand\py[1]{{\lstset{style=python}\lstinline!#1!\lstset{style=python}}}

\usepackage[many]{tcolorbox}    	
\usepackage{setspace}               
\usepackage{multicol}               

\usepackage[numbers,sort&compress]{natbib}

\definecolor{main}{HTML}{5989cf}    
\definecolor{sub}{HTML}{cde4ff}     

\tcbset{
    sharp corners,
    colback = white,
    before skip = 0.2cm,    
    after skip = 0.5cm      
}                           

\newtcolorbox{boxK}{
    sharpish corners, 
    boxrule = 0pt,
    toprule = 4.5pt, 
    enhanced,
    fuzzy shadow = {0pt}{-2pt}{-0.5pt}{0.5pt}{black!35} 
}

\definecolor{codegreen}{rgb}{0,0.6,0}
\definecolor{codegray}{rgb}{0.5,0.5,0.5}
\definecolor{codepurple}{rgb}{0.58,0,0.82}
\definecolor{backcolour}{rgb}{0.95,0.95,0.92}

\lstdefinestyle{mystyle}{
  backgroundcolor=\color{backcolour},   commentstyle=\color{codegreen},
  keywordstyle=\color{magenta},
  numberstyle=\tiny\color{codegray},
  stringstyle=\color{codepurple},
  basicstyle=\ttfamily\footnotesize,
  breakatwhitespace=false,         
  breaklines=true,                 
  captionpos=b,                    
  keepspaces=true,                 
  numbers=left,                    
  numbersep=5pt,                  
  showspaces=false,                
  showstringspaces=false,
  showtabs=false,                  
  tabsize=2
}

\begin{document}

\noindent \today
\hfill 

\vskip 0.4cm

\begin{center}
\bigskip
{\huge\bf
\begin{spacing}{1.1}
ELENA :\\ a software for fast and precise computation of first order phase transitions and gravitational waves production in particle physics models
\end{spacing}
}

\vspace{1.2cm}

\renewcommand*{\thefootnote}{\fnsymbol{footnote}}
{\large
\noindent Francesco Costa$^{1}$, Jaime Hoefken Zink$^{2}$, Michele Lucente$^{3,4}$,\\ Silvia Pascoli$^{3,4}$ and Salvador Rosauro-Alcaraz$^{4}$
\\[3mm]
{\it{
$^1$ Institute of Particle and Nuclear Physics, Faculty of Mathematics and Physics,
Charles University in Prague, V Hole\v{s}ovi\v{c}k\'ach 2, 180 00 Praha 8, Czech Republic \\
$^2$ National Centre for Nuclear Research, Pasteura 7, Warsaw, PL-02-093, Poland \\
$^3$ Dipartimento di Fisica e Astronomia, Universit\`a di Bologna,\\ via Irnerio 46, 40126 Bologna, Italy \\
$^4$ INFN, Sezione di Bologna, viale Berti Pichat 6/2, 40127 Bologna, Italy\\ }
}}
\vspace{0.8cm}
{francesco.costa@matfyz.cuni.cz, 
jaime.hoefkenzink@ncbj.gov.pl, 
michele.lucente@unibo.it, \\
silvia.pascoli@unibo.it,
rosauroa@bo.infn.it}
\end{center}
\vskip 0.4cm

\renewcommand*{\thefootnote}{\arabic{footnote}}
\setcounter{footnote}{0}

\begin{abstract}
We present \elena (EvaLuator of tunnElliNg Actions), an open-source Python package designed to compute the full evolution of first-order phase transitions in the early Universe generated by particle physics models, taking into account several refinements that go beyond commonly assumed simplifications. The core of \elena is based on a vectorized implementation of the tunnelling potential formalism, which allows for a fast computation of the finite-temperature tunnelling action. This, in turn, enables the sampling of the full range of temperatures where two phases coexist and the use of integral expressions that track the complete evolution of the transition, providing a comprehensive picture of it. In addition, \elena provides all the tools to compute the resulting stochastic gravitational waves spectrum, allowing for the full chain of computations -- from the Lagrangian parameter inputs to the final gravitational waves spectrum -- in a fast and self-contained implementation. 
\end{abstract}

\vspace{1cm}
\textbf{Code availability:} \elena is available at \url{https://github.com/michelelucente/ELENA}.

\thispagestyle{empty}

\pagebreak

\begin{small}
\tableofcontents
\end{small}

\newpage


\section{Introduction}

First-order phase transitions (FOPTs) in the early Universe provide a unique window into physics beyond the Standard Model (BSM). FOPTs are both theoretically motivated by many Standard Model (SM) extensions and observationally compelling, given that they can source a detectable stochastic gravitational wave background (SGWB). Indeed, strong FOPTs can be observed by future interferometers such as LISA, Einstein Telescope, Taiji, and Tian. Moreover, recent observation of the Hellings-Downs angular correlation pattern in multiple Pulsar Timing Array (PTA) datasets~\cite{NANOGrav:2023gor,EPTA:2023fyk,Reardon:2023gzh,Xu:2023wog}, including those from NANOGrav, EPTA, PPTA, and IPTA, has provided the first evidence for the existence of a SGWB in the nHz frequency range. Although binary systems of supermassive black holes (SMBHs) provide a plausible astrophysical interpretation of the observed signal ~\cite{1995ApJ...446..543R,Wyithe_2003,Jaffe_2003}, cosmological first-order phase transitions have emerged as one of the leading theoretical scenarios, offering a well-motivated cosmological origin for the SGWB and a better fit to the data~\cite{Gouttenoire:2023bqy, Chen:2023bms, Wang:2023bbc, Ghosh:2023aum, Addazi:2023jvg, Bringmann:2023opz, Croon:2024mde, Winkler:2024olr, Conaci:2024tlc, Fujikura:2023lkn, Banik:2024zwj}.

A first-order phase transition arises when a scalar field acquires a non-zero vacuum expectation value. It proceeds via the nucleation of bubbles of a true vacuum within a metastable background, triggered by quantum or thermal fluctuations. The dynamics of bubble nucleation and expansion, the profile of the effective potential at finite temperature, and the tunnelling action all play crucial roles in determining the evolution of the transition and its observable consequences. While existing tools like \textsf{CosmoTransitions}~\cite{Wainwright:2011kj}, \textsf{BSMPT}~\cite{Basler:2024aaf}, \textsf{TransitionListener}~\cite{Ertas:2021xeh}, \textsf{PT2GWFinder}~\cite{Brdar:2025gyo}, and \textsf{PhaseTracer}~\cite{Athron:2024xrh} can tackle some or all of these problems, they all rely on the bounce action method, which can be slow or numerically unstable for some regions of the model parameter space.

We present \elena (EvaLuator of tunnElliNg Actions), a new open-source Python package designed to provide fast and precise computations for FOPTs induced by scalar field potentials from particle physics models. \elena builds upon the tunnelling potential formalism~\cite{Espinosa:2018hue, Espinosa:2018szu}, an alternative to the traditional bounce method that recasts the vacuum decay problem as a minimization task, converting an equation of motion problem into an optimization one. This formulation provides significant numerical advantages: it is faster and more stable, and therefore is naturally suited for scans over parameter spaces, which are often required in phenomenological studies of new physics.

In addition to computing the finite-temperature tunnelling action, \elena tracks the full evolution of the phase transition, determining key-milestone temperatures (namely critical, nucleation, percolation, and completion temperatures), and computing the thermal parameters relevant to GW generation: the strength and speed of the transition, the mean bubble separation, or the plasma sound speed, among others. The package provides a complete pipeline from Lagrangian inputs to GW spectra, making it a practical and powerful tool for model building and phenomenology.

\elena provides ready-to-run examples and accepts potentials generated from the widely used \textsf{CosmoTransitions} model class, thus ensuring back-compatibility with a large number of already implemented models. 

\elena thus complements and enhances the current ensemble of computational tools in the field, contributing to the broader effort to probe early Universe physics via upcoming GW observatories.

\section{The tunnelling potential formalism}\label{sec:formalism}
\begin{figure}[htb]
    \centering
    \includegraphics[width=0.7\linewidth]{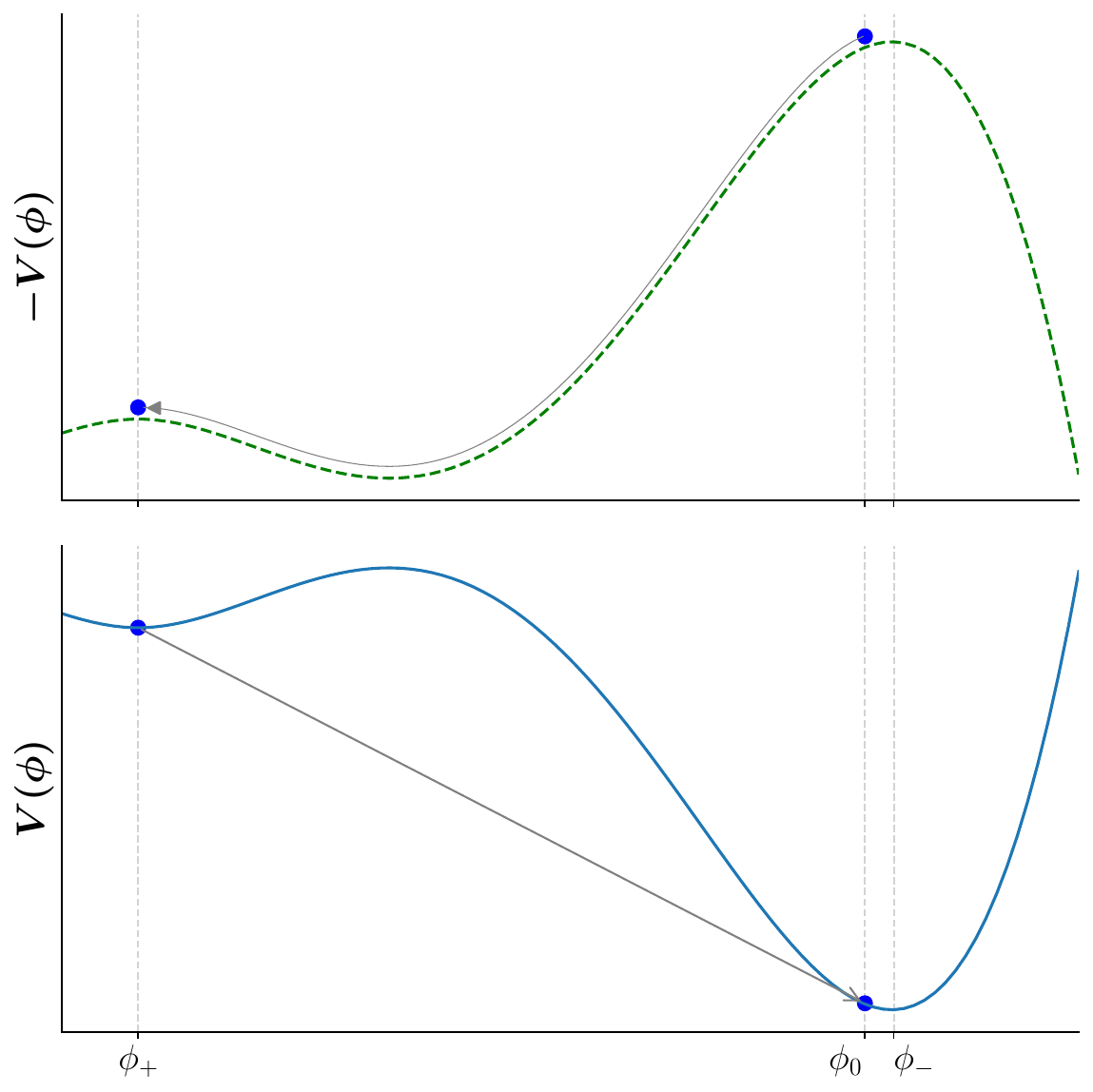}
    \caption{
    Pictorial representation of the classical mechanical dual of the bounce equations of motion,~\Cref{eq:bounce,eq:boundary}. \emph{Upper panel:} a classical particle at position $\phi$ is subject to the potential $-V(\phi)$, with a friction term that decreases with time $\rho$ (see~\Cref{eq:bounce}). The solution satisfying the boundary conditions in~\Cref{eq:boundary} amounts to finding the value of $\phi_0$ such that the particle, starting at rest from $\phi = \phi_0$ at $\rho = 0$, stops at $\phi = \phi_+$ for $\rho \rightarrow \infty$.
    \emph{Lower panel:} the scalar field nucleates via an $O(d)$-symmetric field configuration, with value $\phi_0$ at $\rho = 0$ and $\phi_+$ at $\rho \rightarrow \infty$, where $\rho$ is the Euclidean radius in $d$ dimensions. The field profile at finite values of $\rho$ is determined by the solution $\phi_b(\rho)$ to~\Cref{eq:bounce,eq:boundary}.
    }
    \label{fig:bounce}
\end{figure}

The core module of \elena is the implementation of the tunnelling potential formalism~\cite{Espinosa:2018hue}~\footnote{This formalism has also been extended to multi-field potentials~\cite{Espinosa:2018szu}, although \elena focuses on one-field potentials in its first version.}, which is used to compute the tunnelling action for the decay of a scalar field $\phi$ from a metastable false vacuum state $\phi_+$, to a true vacuum state $\phi_-$.
In the seminal work~\cite{Coleman:1977py}, it was demonstrated that, under the assumption that solutions are $O(d)$-symmetric and given a potential $V(\phi)$, the field configuration $\phi_b(\rho)$ that interpolates between $\phi_+$ and the basin of $\phi_-$ obeys the so-called `bounce equation'
\begin{equation}\label{eq:bounce}
    \ddot{\phi} (\rho) + \frac{d-1}{\rho} \dot{\phi}(\rho) = V'(\phi)\,,
\end{equation}
where the dot and prime represent the derivative with respect to $\rho$ and $\phi$, respectively, and $\rho = \sqrt{\tau^2 + |\vec{x}|^2}$ is the Euclidean radius in $d$ dimensions (with $\tau$ the Euclidean time).~\Cref{eq:bounce} is complemented by the boundary conditions
\begin{equation}\label{eq:boundary}
    \dot{\phi_b} (0) = 0, \hspace{2cm} \lim_{\rho \rightarrow \infty} \phi_b (\rho) = \phi_+,
\end{equation}
which ensures that the solution remains non-singular at the origin and that the field remains in the false vacuum state far from the nucleated bubble. Once the solution $\phi_b(\rho)$ is known, the tunnelling action can be determined by inserting it into the $d$-dimensional Euclidean action
\begin{equation}
    S_{E,d} = \frac{2 \pi^{d/2}}{\Gamma(d/2)} \int_0^\infty \left[\frac{1}{2} \dot{\phi}_b^2 + V\left(\phi_b\right) - V(\phi_+) \right] \rho^{d-1} \de \rho. 
\end{equation}
\Cref{eq:bounce} with the boundary conditions in~\Cref{eq:boundary} has an intuitive classical mechanics interpretation: by identifying $\phi$ as the position of a classical particle subject to a potential $-V$ and to a friction term that decreases over time as $\rho^{-1}$, with $\rho$ being the time variable, the solution to the problem is equivalent to finding the position $\phi_0$ such that the particle, starting at rest from $\phi = \phi_0$ at $\rho = 0$, rolls down and exactly stops at $\phi = \phi_+$. It is clear from~\Cref{fig:bounce} that the solution to this problem is an unstable point~\cite{Coleman:1977py}: values of $\phi$ slightly larger than $\phi_0$ will result in the particle overshooting the point $\phi_+$ and rolling away from it, while slightly smaller values make the particle roll back and oscillate around the local minimum of $-V$ (corresponding to the maximum of $V$). This observation is the basis for a popular numerical recipe~\cite{Wainwright:2011kj} to solve~\Cref{eq:bounce,eq:boundary}, known as the overshoot-undershoot method: it amounts to solving the equations of motion for a trial value of $\phi_0$. Depending on the asymptotic behaviour of the field value (that is, depending on whether the trajectory undershoots or overshoots $\phi_+$), the initial position $\phi_0$ is displaced closer or further from the top. This procedure is iterated until the required numerical precision is reached. 

The undershoot-overshoot method can be numerically demanding, given that small deviations from the correct value of $\phi_0$  result in large differences of the asymptotic field value from $\phi_+$. In addition, the convergence time of the solution tends to infinity for infinitesimal energy differences between the false and true vacua, although analytical solutions for this regime are known~\cite{Coleman:1977py}.

The tunnelling potential formalism~\cite{Espinosa:2018hue} addresses the numerical issues of the bounce formalism by redefining the equations of motion, such that the correct value of $\phi_0$ becomes a minimum of the so-called tunnelling action. 
The numerical advantage is twofold: first and foremost, finding the minimum of a function is much easier than finding a saddle point. Moreover, it is possible to completely remove the dependency on the bounce equations~\Cref{eq:bounce,eq:boundary}, so that no differential equation needs to be numerically integrated.

The tunnelling potential is an auxiliary function defined as
\begin{equation}
    V_t (\phi) \equiv V(\phi) - \frac{1}{2} \dot{\phi}_b^2,
\end{equation}
where $\phi_b$ is the solution to the bounce equations (\Cref{eq:bounce,eq:boundary}). As demonstrated in Ref.~\cite{Espinosa:2018hue}, $V_t(\phi)$ is a monotonic function with $V_t(\phi) \leq V(\phi)$, and is defined only in the range $\phi \in [\phi_+, \phi_0]$, with $V_t = V$ at the endpoints. As anticipated, it is possible to remove the dependence on the bounce solution $\phi_b$ by employing the relation
\begin{equation}
    \dot{\phi_b} = - \sqrt{2\left[V(\phi) - V_t(\phi) \right]},
\end{equation}
as well as on the radial variable by using
\begin{equation}
    \rho = \left( d -1 \right) \sqrt{\frac{V - V_t}{\left( V_t'\right)^2}}.
\end{equation}

The new equations of motion are given by
\begin{equation}\label{eq:tunneling}
    \left( V_t' \right)^2 = \frac{d-1}{d} \left[ V' V_t' - 2 \left(V_t - V\right) V_t''\right],
\end{equation}
with boundary conditions
\begin{equation}\label{eq:tunnelling_boundary}
    V_t\left(\phi_+\right) = V\left(\phi_+\right),\hspace{1cm}  V_t\left(\phi_0\right) = V\left(\phi_0\right).
\end{equation}
The new problem is to find the correct value of $\phi_0$ and the form of $V_t(\phi)$ that solves~\Cref{eq:tunneling,eq:tunnelling_boundary}. Although this does not seem an easier task with respect to the starting point in~\Cref{eq:bounce,eq:boundary}, as demonstrated in Ref.~\cite{Espinosa:2018hue}, one does not need to solve this differential equation in practice. An excellent approximation for $V_t$ can indeed be derived by considering the shape of the scalar potential $V$; together with the observation that $\phi_0$ is a minimum of the tunnelling action $S_{E,d}$, given by
\begin{equation}\label{eq:action}
    S_{E,d} = \frac{\left(d-1\right)^{(d-1)} \left(2\pi \right)^{\frac{d}{2}}}{\Gamma\left(1 + \frac{d}{2} \right)} \int_{\phi_+}^{\phi_0} \frac{\left(V - V_t\right)^\frac{d}{2}}{\left|V_t'\right|^{(d-1)}} \de \phi,
\end{equation}
this allows to apply the following numerical algorithm to solve the problem:

\begin{boxK}
Construct an approximation to $V_t(\phi)$ following the procedure outlined in Ref.~\cite{Espinosa:2018hue}, and then insert it in~\Cref{eq:action}, computing $S_{E,d}$ as a function of $\phi_0$. The solution to the problem is given by the value of $\phi_0$ that minimises $S_{E,d}$.
\end{boxK}

This is exactly the numerical procedure implemented in \elena, contained in the class \py{espinosa.Vt_vec} and described in more detail in the following Sections.

\begin{figure}[htb]
    \centering
    \includegraphics[width=0.99\linewidth]{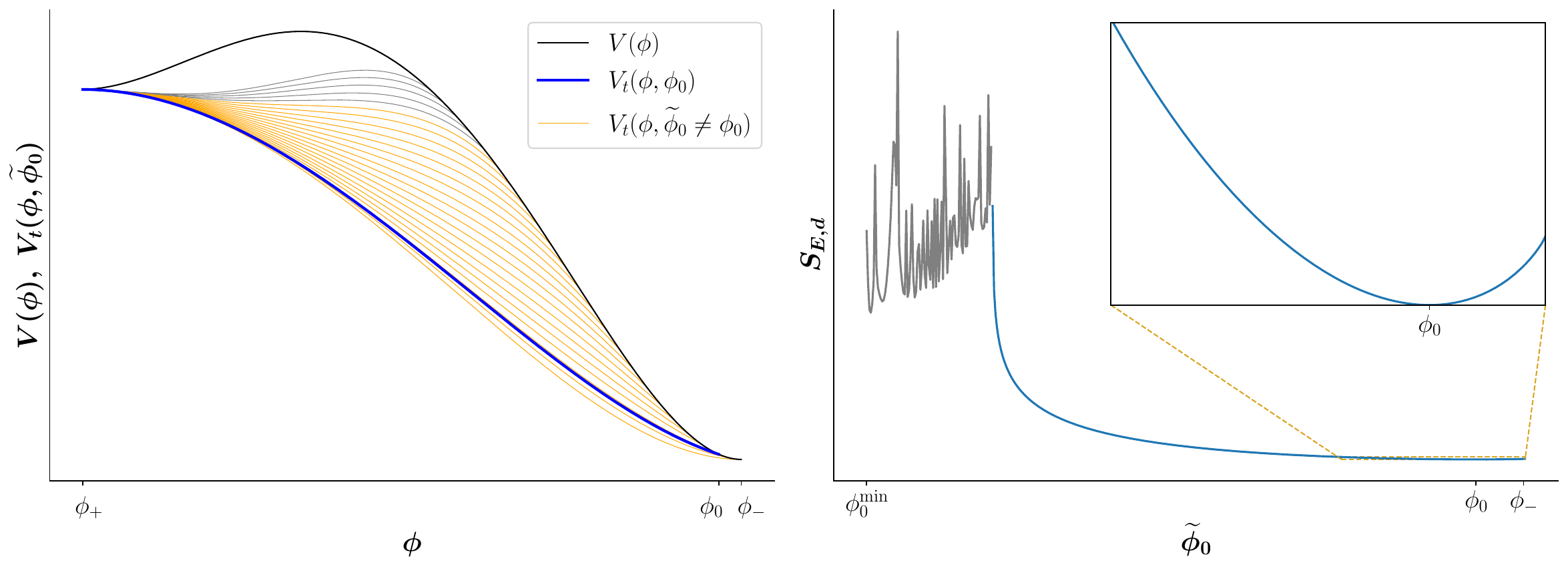}
    \caption{
    Example of the tunnelling potential formalism. 
    \emph{Left panel:} the tunnelling potential $V_t(\phi, \widetilde{\phi}_0)$ for different values of the $\widetilde{\phi}_0$ parameter, obtained by following the procedure outlined in Ref.~\cite{Espinosa:2018hue}; the blue line indicates the potential $V_t(\phi, \phi_0)$, obtained by choosing the value of $\widetilde{\phi}_0= \phi_0$ that minimises the action $S_{E,d} (\widetilde{\phi}_0)$. Gray lines represent configurations where the construction of the tunnelling potential breaks down, while orange lines are numerically viable configurations. For reference, the potential $V(\phi)$ is plotted as a black line.
    \emph{Right panel:} the value of the action $S_{E,d}(\widetilde{\phi}_0)$ as a function of the $\widetilde{\phi}_0$ parameter, computed following~\Cref{eq:action}. $\phi_0^\textrm{min}$ is the smallest value of $\widetilde{\phi}_0$ for which the tunnelling can take place, that is, the smallest field value in the region to the right of the local barrier for which $V(\phi) < V(\phi_+)$.  Values in gray correspond to unreliable $V_t$ constructions (cf. the gray lines in the left panel), while the blue line corresponds to numerically reliable computations. The inset box magnifies the curve around $\widetilde{\phi}_0 = \phi_0$, the value of $\widetilde{\phi}_0$ that minimises $S_{E,d}$.
    }
    \label{fig:tunnelling}
\end{figure}

In~\Cref{fig:tunnelling} we report an example of the formalism. In the left panel, we show the potential $V(\phi)$ (black line) together with the tunnelling potential $V_t(\phi, \widetilde{\phi}_0)$. This notation means that $V_t$ is constructed following the prescription in Ref.~\cite{Espinosa:2018hue} for an arbitrary value $\widetilde{\phi}_0$ of the field at the centre of the nucleated bubble, which is not necessarily the solution to the equations of motion. The correct form for $V_t$, which is the one constructed by choosing the value $\widetilde{\phi}_0 = \phi_0$ minimising the tunnelling action, is shown as a blue line. We notice that, for values of $\widetilde{\phi}_0$ that are excessively far-away from the correct one, the construction in Ref.~\cite{Espinosa:2018hue} breaks down as $V_t$ is no longer monotonically decreasing; we identify these problematic tunnelling potentials by gray lines, while the well-behaved ones are identified by orange lines. This behaviour is nevertheless not an issue for the numerical routine, since the action $S_{E,d}(\widetilde{\phi}_0$) sharply increases when moving towards this problematic region (see right panel in~\Cref{fig:tunnelling}) such that the results of the calculation within the region itself systematically give values of the tunnelling action that are orders of magnitude larger than the minimum of $S_{E,d}$ at $\phi_0$ (even if the result of the computation cannot be trusted). We conclude that, since \elena looks for the minimum of $S_{E,d}$ as a function of $\widetilde{\phi}_0$, the results are reliable: the algorithm works well in the region of interest, while in regions where the method is unreliable, the computed value of $S_{E,d}$ remains substantially larger than its minimum, preventing these configurations from being misidentified as valid solutions to the equations of motion.

\section{From the Lagrangian model parameters to the gravitational wave spectrum using \elena}\label{sec:software_chain}

In this Section we outline how to use \elena to perform the full chain of computations necessary to predict the SGWB from a first order phase transition, inputting only the model Lagrangian parameters. In each step, we describe the relevant modules used by \elena. The full code is available as a ready-to-run \textsf{Jupyter} notebook in \py{./examples/phase_transition.ipynb}. We summarise in Table~\ref{tab:comparison_codes} \elena's main features, and compare with those from other popular codes~\cite{Wainwright:2011kj,Athron:2019nbd,Athron:2020sbe,Athron:2024xrh,Basler:2024aaf,Guada:2020xnz,Brdar:2025gyo}. 

\begin{table}
    \centering
    \begin{tabular}{| W{c}{3.5cm} || W{c}{1cm} | W{c}{3.0cm} | W{c}{0.5cm} | W{c}{0.5cm} | W{c}{0.5cm} | W{c}{1cm} | W{c}{3.2cm}|}
    \hline
    Software & Phases & $S_3/T$ \& Method & $T_n$ & $T_p$ & $T_e$ & GW & Assumptions\\ 
    \hline
    \hline
    \textcolor{blue}{\elena} & \CheckmarkBold & \begin{tabular}{@{}c@{}c@{}c@{}}\CheckmarkBold\\ Single field\\ Tunneling\\ potential\end{tabular} & \CheckmarkBold & \CheckmarkBold & \CheckmarkBold & \CheckmarkBold & \begin{tabular}{@{}c@{}}Adiabatic\\ expansion\end{tabular}\\
    \hline
    \textcolor{blue}{\textsf{CosmoTransitions}}~\cite{Wainwright:2011kj} & \CheckmarkBold & \begin{tabular}{@{}c@{}c@{}c@{}}\CheckmarkBold\\ Multi-field\\Shooting and\\ path deformation\end{tabular} & \CheckmarkBold & \XSolidBrush & \XSolidBrush & \XSolidBrush & $S_{E,3}(T_n)/T_n=140$\\
    \hline
    \textcolor{blue}{\textsf{TransitionListener}}~\cite{Ertas:2021xeh} & \CheckmarkBold & \begin{tabular}{@{}c@{}c@{}c@{}}\CheckmarkBold\\ Multi-field\\Shooting and\\ path deformation\end{tabular} & \CheckmarkBold & \XSolidBrush & \XSolidBrush & \CheckmarkBold & \begin{tabular}{@{}c@{}c@{}c@{}c@{}c@{}c@{}}Adiabatic\\ expansion\\$H(T)^2\propto \rho_{\mathrm{rad}}(T)$\\$\Gamma(T_n)/H^4(T_n)=1$ \\ $\Gamma(t)\sim \Gamma_ne^{\beta t}$ \\ GW at $T_n$
    \end{tabular}\\
    \hline
    \textcolor{orange}{\textsf{BubbleProfiler}}~\cite{Athron:2019nbd} & \XSolidBrush & \begin{tabular}{@{}c@{}c@{}c@{}}\CheckmarkBold\\Multi-field\\ Shooting and\\ perturbative\end{tabular} & \XSolidBrush & \XSolidBrush & \XSolidBrush & \XSolidBrush & None\\
    \hline
    \textcolor{orange}{\textsf{PhaseTracer2}}~\cite{Athron:2024xrh} & \CheckmarkBold & \begin{tabular}{@{}c@{}c@{}c@{}}\CheckmarkBold\\Multi-field\\ Shooting and\\ path deformation\end{tabular} & \CheckmarkBold & \XSolidBrush & \XSolidBrush & \CheckmarkBold & \begin{tabular}{@{}c@{}c@{}c@{}c@{}c@{}c@{}}Adiabatic\\ expansion\\$H(T)^2\propto \rho_{\mathrm{rad}}(T)$\\ $S_{E,3}(T_n)/T_n=140$\\ $\Gamma(t)\sim \Gamma_ne^{\beta t}$\\ GW at $T_n$\end{tabular} \\
    \hline
    \textcolor{orange}{\textsf{BSMPT}}~\cite{Basler:2024aaf} & \CheckmarkBold & \begin{tabular}{@{}c@{}c@{}c@{}}\CheckmarkBold\\Multi-field\\ Shooting and\\ path deformation\end{tabular} & \CheckmarkBold & \CheckmarkBold & \CheckmarkBold & \CheckmarkBold & \begin{tabular}{@{}c@{}c@{}c@{}c@{}c@{}c@{}c@{}}Adiabatic\\expansion\\$H(T)^2\propto \rho_{\mathrm{rad}}(T)$\\Constant $g_*$\\MIT bag model\\$\Gamma(t)\sim \Gamma_ne^{\beta t}$\\GW from sound \\waves \& turbulence\end{tabular} \\
    \hline
    \textcolor{violet}{\textsf{FindBounce}}~\cite{Guada:2020xnz} & \XSolidBrush & \begin{tabular}{@{}c@{}c@{}c@{}}\CheckmarkBold\\Multi-field\\ Polygonal  bounces\end{tabular} & \XSolidBrush & \XSolidBrush & \XSolidBrush & \XSolidBrush & None\\
    \hline
    \textcolor{violet}{\textsf{PT2GWFinder}}~\cite{Brdar:2025gyo} & \CheckmarkBold & \begin{tabular}{@{}c@{}c@{}c@{}}\CheckmarkBold\\ Single field\\Polygonal bounces\end{tabular} & \CheckmarkBold & \CheckmarkBold & \CheckmarkBold & \CheckmarkBold & \begin{tabular}{@{}c@{}c@{}c@{}c@{}c@{}}Adiabatic\\expansion\\Constant $g_*$\\MIT bag model\\$\Gamma(t)\sim \Gamma_ne^{\beta t}$\end{tabular} \\
    \hline
    \end{tabular}
    \caption{Comparison of the features included in \elena and other popular codes, along with the approximations adopted in their pipelines. We use different colors for the name of each software to highlight the language in which they are coded. We use \textcolor{blue}{blue} for \textsf{Python}, \textcolor{orange}{orange} for \textsf{C++}, and \textcolor{violet}{violet} for \textsf{Mathematica}. We note that both \textsf{PT2GWFinder} and \textsf{PhaseTracer2} can be used together with \textsf{DRAlgo}~\cite{Ekstedt:2022bff}. Moreover, while \textsf{PhaseTracer2} uses the nucleation temperature to compute the GW spectrum, we point out the existence of \textsf{TransitionSolver}, developed by the same authors, which can, in principle, compute all parameters. Given that the latter is still in developing stages, we refrain from comparing to it. \textsf{TransitionListener} uses \textsf{CosmoTransitions} to find the phases and to compute the action, adding GW-related features to it.}
    \label{tab:comparison_codes}
\end{table}

\subsection{Construction of the finite temperature scalar potential}
\elena uses the \py{model} class from \CT~\cite{Wainwright:2011kj} to construct the finite-temperature scalar potential. We thus refer the user to the \CT documentation for details on the implementation of a new model. Once the potential has been constructed, \elena uses its own modules for the computation of the phase transition dynamics, not relying on \CT any longer.

The current version of \elena includes the implementation of the dark sector model presented in Ref.~\cite{Costa:2025csj}, composed of a complex scalar $\phi$, charged under a new $U(1)_D$ gauge symmetry, and its associated dark photon $Z_\mu^\prime$. Its Lagrangian is given by
\begin{equation}
\mathcal{L}= \left(D_{\mu}\phi\right)^{*}\left(D^{\mu}\phi\right) -\frac{1}{4}Z^{\prime}_{\mu\nu}Z^{\prime \mu\nu} -V(\phi^*\phi),\hspace{1cm} V=-\mu_{\phi}^2\phi^*\phi+\lambda_{\phi}\left(\phi^*\phi\right)^2,
\label{eq:lag_UV}
\end{equation}
where $D_{\mu}\equiv \partial_{\mu}-i\sqrt{2}g_D Z^{\prime}_{\mu}$ is the covariant derivative and $g_D$ the dark gauge coupling.
The potential in~\Cref{eq:lag_UV} with $\mu_\phi^2, \lambda_\phi > 0$ has a global minimum at $v_\phi^0=\mu_{\phi} / \sqrt{\lambda_\phi}$; when the scalar field acquires a vacuum expectation value (vev) $v_\phi^0$, the $U(1)_D$ symmetry is spontaneously broken, resulting in the mass spectrum $m_{Z'}^2 = g_D^2 (v_\phi^0)^2$ and $m_{\phi}^2 = 2\lambda_{\phi} (v_\phi^0)^2$.

The implementation of the potential in~\Cref{eq:lag_UV} provided in \elena employs the on-shell renormalisation method, which is expected to give more conservative predictions for the resulting SGWB~\cite{Zhu:2025pht}. Consequently, the quantities appearing in the Lagrangian match the observable physical ones. 

Once a model class has been implemented, the finite temperature potential is created by simply initialising an instance of the class with the desired Lagrangian parameters, as shown in the~\Cref{code:model}.

\begin{lstlisting}[language=Python, caption=Finite temperature potential implementation and initialization in \elena., label=code:model]
from model import model

# Select the value of the model parameters
lambda_ = 1.65e-3
g = 0.54
vev = 500

# This constructs the finite-temperature potential as an instance of the class model
dp = model(vev, lambda_, g, xstep = 1e-3 * vev, Tstep = 1e-3 * vev) # Dark photon model instance
V = dp.DVtot # Scalar potential
dV = dp.gradV # Gradient of the scalar potential
\end{lstlisting}

The arguments \py{xstep} and \py{Tstep} are internal arguments of \CT. We find it useful to normalise them to the energy scale of the problem (i.e. the vev value) to improve the numerical stability of the computation.

In the following, we present two examples of a phase transition computation for different choices of the model parameters: the first one is dubbed ``Fast'', given that the transition takes place on a timescale much smaller than the Hubble time; in the second one, dubbed ``Slow'', the duration of the transition is instead comparable to it (cf.~\Cref{sec:bubble_radius} for numerical values).
The input Lagrangian parameters for each case are collected in~\Cref{tab:input}.
\begin{table}[htb]
    \centering
    \begin{tabular}{|c|c|c|}
    \hline
        Parameter & Fast & Slow \\
        \hline
         $\lambda_\phi$ & $1.65 \cdot 10^{-3}$ & $6 \cdot 10^{-3}$ \\
         $g_D$ & $0.54$ & $0.75002$ \\
         $v_\phi^0$ & $500$ MeV   & $500$ MeV \\
    \hline
    \end{tabular}
    \caption{Input parameters for the two example benchmark points analysed in this work, as referred to the Lagrangian from~\Cref{eq:lag_UV}. The ``Slow'' point coincides with the BP1 in Ref.~\cite{Costa:2025csj}.
    }
    \label{tab:input}
\end{table}

\subsection{Computation of the critical temperatures}
A FOPT can only happen if a barrier separating the true and false vacua is present. This condition determines the range of temperatures $T \in [T_{min}, T_c]$ for which a FOPT is possible: $T_c$ is the critical temperature, defined as the temperature at which the values of the potential in the true and false vacua are degenerate and a potential barrier separates the two minima; $T_{min}$ is the temperature below which a potential barrier is no longer present.
For $T>T_c$, the global minimum of the potential is the one corresponding to the false vacuum; for $T< T_{min}$ the field can simply roll to the true minimum, a process that does not result in bubble nucleation.

\elena implements its own routines, \py{temperatures.find_T_min} and \py{temperatures.find_T_max}, to compute the temperature range in which the FOPT is possible. The algorithm looks for changes in the sign of the gradient of the potential to find the local extrema of the potential itself. Next, it checks if a FOPT is possible given the shape of the potential at a given temperature. The routines \py{temperatures.find_T_min} and \py{temperatures.find_T_max} loop over different temperature values (starting from small and large temperatures, respectively) to verify the smallest and largest temperatures for which a FOPT can take place, reducing the size of the step at each iteration until a desired precision is met.
Once a first estimation from \py{temperatures.find_T_min} is known, \elena provides the function \py{temperatures.refine_Tmin} to further refine the determination of $T_{min}$ by looping over a narrower range of temperatures using smaller steps. This allows to obtain an accurate result without a significant numerical overload. Such a precise determination can be necessary for very long transitions, when the completion temperature is close to $T_{min}$. We do not implement an analogous function to refine $T_c$ because the nucleation probability is infinitesimal at $T_c$, and we find that the computation of the transition dynamics does not change appreciably with a more precise determination of $T_c$ with respect to the one already obtained by \py{temperatures.find_T_max}.
An example computation of the critical temperatures using \elena is shown in the~\Cref{code:critical_T}.

\begin{lstlisting}[language=Python, caption=Computation of the critical and minimum temperatures., label=code:critical_T]
from temperatures import find_T_min, find_T_max, refine_Tmin

T_max, vevs_max, max_min_vals, false_min_tmax = find_T_max(V, dV, precision = 1e-2, Phimax = 2 * vev, step_phi = vev * 1e-2, tmax = 2.5 * vev)
T_min, vevs_min, false_min_tmin = find_T_min(V, dV, tmax = T_max, precision = 1e-2, Phimax = 2 * vev, step_phi = vev * 1e-2, max_min_vals = max_min_vals)

if T_max is not None and T_min is not None:
    maxvev = np.max(np.concatenate((vevs_max, vevs_min)))
elif T_max is not None:
    maxvev = np.max(vevs_max)
elif T_min is not None:
    maxvev = np.max(vevs_min)
else:
    maxvev = None

T_min = refine_Tmin(T_min, V, dV, maxvev, log_10_precision = 6) if T_min is not None else None
\end{lstlisting}

Apart from the determination of $T_c$ (\py{T_max}), the function \py{temperatures.find_T_max} also returns other quantities that are useful to improve the convergence of subsequent computations: a list of vev values at finite temperature (\py{vevs_max}), an array containing the field values where the potential $V(\phi, T_c)$ is at its global maximum and minimum (\py{max_min_vals}, corresponding to the locations of the barrier maximum and true minimum at $T_c$), and an array containing the field location and potential value at the false minimum at $T_c$ (\py{false_min_tmax}). Analogously, \py{temperatures.find_T_min} returns the value of $T_{min}$ and two auxiliary quantities, here saved as \py{vevs_min} and \py{false_min_tmin}.
Optional arguments to the above described functions are: \py{precision} (default = \py{1e-2}) that sets the desired precision of the computation\footnote{The loops stop when the ratio between the height of the potential barrier and the depth of the true minimum is smaller than \py{precision} for \py{temperatures.find_T_min}, or when the difference between the potential value at true and false minima divided by the height of the barrier is smaller than \py{precision} for \py{temperatures.find_T_max}.}, \py{Phimax} (set by default to \py{150}) that sets the maximum value of the field to be considered, \py{step_phi} (default \py{0.1}) that sets the step size in the field dimension when constructing the potential and gradient, and \py{tmax} and \py{tmin} (default \py{250} and \py{0}), which set the maximal and minimal temperatures to be considered in the search.~\footnote{Given the scale invariance of the potential, we do not specify the units on these dimensionful variables, but note that these are assumed to be the same between field and temperature variables and set by \py{units}.} The optional argument \py{max_min_vals}, only present in \py{temperatures.find_T_min}, can improve the computation by providing information on the barrier and true minimum locations.

Once the value of $T_{min}$ has been determined by \py{temperatures.find_T_min}, it can be refined by using \py{temperatures.refine_Tmin}, which checks the minimum value of the form \py{T_min} \py{*} \py{(1 - k} \py{*} \py{10**(-log_10_precision))}, with \py{k} being an integer, for which a barrier is still present. In contrast to the routine in \py{temperatures.find_T_min}, the function \py{temperatures.refine_Tmin} only uses the size of the temperature step as a criterion to stop the refinement, disregarding the height of the barrier. 

\begin{figure}[htb]
    \centering
    \includegraphics[width=0.99\linewidth]{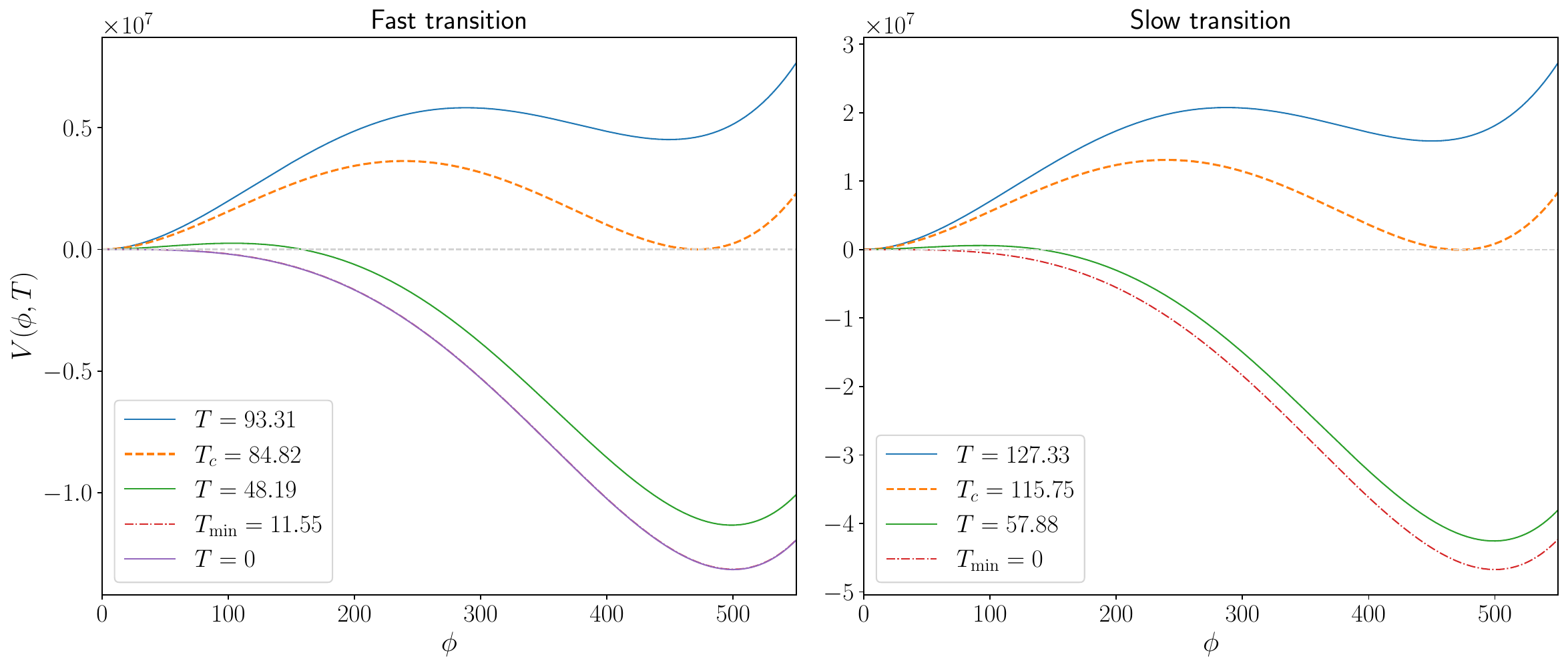}
    \caption{
    Shape of the finite temperature potential for the example points in~\Cref{tab:input} at different temperatures, corresponding to $T = 1.1 \cdot T_c$ (blue line), where $T_c$ is the critical temperature, $T_c$ (orange dashed-line), $T = (T_c - T_{min})/2$ (green line), where $T_{min}$ is the smallest temperature at which a barrier is present, and $T_{min}$ (red dot-dashed line). Additionally, when $T_{min}\neq 0$, we show the potential at $T=0$ as a purple line. Up to a global rescaling of $V$, these results are valid for an arbitrary energy scale (cf.~\Cref{eq:scaling}). The values of the temperatures shown in the legend correspond to the same ones chosen for $\phi$ and given in \protect\py{units}.
    }
    \label{fig:critical_temperatures}
\end{figure}

The finite temperature potential for the points in~\Cref{tab:input} is plotted in~\Cref{fig:critical_temperatures} for different values of the temperature, including $T_c$ (orange dashed line) and $T_{min}$ (red dot-dashed line), together with values at $1.1 * T_c$, $(T_c - T_{min}) / 2$ and $T=0$ for reference (blue, green, and purple lines, respectively). As can be seen, both the height of the barrier and the depth of the true minimum are larger in the ``Slow'' case, resulting in a stronger transition in this scenario.
Moreover, in this case the barrier persists down to zero temperature as a result of the Coleman-Weinberg corrections to the zero temperature potential. The slow nature of the transition is typical of conformal-like potentials (cf. e.g.~\cite{Kierkla:2023von,Sojka:2024btp}). 
We do not indicate the energy scale in~\Cref{fig:critical_temperatures} because the potential enjoys the scaling symmetry
\begin{equation}\label{eq:scaling}
V(\phi, \lambda, g, v, T) = \frac{V(\xi \phi, \lambda, g, \xi v,\xi T)}{\xi^4},
\end{equation}
for arbitrary rescaling $\xi$ of the dimensionful quantities, meaning that up to a global factor in $V$ the plotted shapes are valid for arbitrary choices of the energy scale.

\subsection{Computation of the tunnelling action}
Once the range of temperatures $T \in [T_{min}, T_c]$ over which a FOPT is possible has been determined, the most important quantity to be computed is the tunnelling action in~\Cref{eq:action}, which governs the evolution of the transition. At finite temperature, it enters in the false vacuum decay rate as~\cite{Coleman:1973jx,Coleman:1977py,Linde:1980tt}
\begin{equation}
\Gamma(T)\simeq T^4\left(\frac{S_{E,3}}{2\pi T}\right)^{3/2}e^{-S_{E,3}/T}\,.
\label{eq:nuc_rate}
\end{equation}
The computation of $S_{E,3}$ in \elena is performed using the class \py{espinosa.Vt_vec}, where the name \py{Vt_vec} stands for a vectorised implementation of the tunnelling potential algorithm discussed in~\Cref{sec:formalism}. This class computes, for a given potential, the relevant quantities appearing in~\Cref{eq:action} using a discrete grid of values for the $\phi$ and $\widetilde{\phi}_0$ variables, and selects as solution the value of $\widetilde{\phi}_0 = \phi_0$ that minimises $S_{E,3}$. The step size \py{step_phi} over which to scan $\phi$ is chosen by the user, while the step size over $\widetilde{\phi}_0$ starts with a relatively large value exploring the full range of possible solutions; at subsequent iterations, the target space of $\widetilde{\phi}_0$ is reduced around an interval containing $\phi_0$ while the step size is decreased correspondingly. The iterations continue until the relative difference in the value of the action computed in two subsequent iterations is smaller than a user-defined value given by \py{precision}.

The discretised computation performed by \py{espinosa.Vt_vec} employs fully vectorised {\textsf{NumPy}\xspace} arrays~\cite{harris2020array}, allowing for a numerically efficient computation. Notice that no numerical differential equation needs to be solved to determine $\phi_0$, thus allowing for a fast computation.

The computation of the tunnelling action in \elena is done by creating an instance of the class \py{espinosa.Vt_vec} with the desired model parameters, and then extracting the attribute \py{action_over_T} from that instance. We provide an example code in the~\Cref{code:action_over_T}, while stressing that \py{espinosa.Vt_vec} can be implemented in more general coding structures according to the specific needs of the user. In addition to \py{action_over_T}, an instance of \py{espinosa.Vt_vec} contains several other attributes that provide further quantities relevant to the tunnelling process. These are exemplified as well in the~\Cref{code:action_over_T}.

\begin{lstlisting}[language=Python, caption=Example of implementation of a tunnelling action computation., label=code:action_over_T]
import numpy as np
from espinosa import Vt_vec

true_vev = {}
S3overT = {}
V_min_value = {}
phi0_min = {}
V_exit = {}
false_vev = {}

def action_over_T(T, c_step_phi = 1e-3, precision = 1e-3):
    instance = Vt_vec(T, V, dV, step_phi = c_step_phi, precision = precision, vev0 = maxvev, int_threshold = 2e-1)
    if instance.barrier:
        true_vev[T] = instance.true_min
        false_vev[T] = instance.phi_original_false_vev
        S3overT[T] = instance.action_over_T
        V_min_value[T] = instance.min_V
        phi0_min[T] = instance.phi0_min
        V_exit[T] = instance.V_exit
        return instance.action_over_T
    else:
        return None

n_points = 100
temperatures = np.linspace(T_min, T_max, n_points)
action_vec = np.vectorize(action_over_T)
action_vec(temperatures)
\end{lstlisting}

Several \python dictionaries are created in the~\Cref{code:action_over_T} to store the result of the computation for different values of the temperature parameter. A function \py{action_over_T()} is then defined as a wrapper of \py{espinosa.Vt_vec}, to compute an instance of the class for a specific temperature and store the computed attributes in the dictionaries. Useful attributes of an instance of the class include:
\begin{itemize}
    \item \py{action}: the value of the finite-temperature $d$-dimensional tunnelling action $S_{E,d}$;
    \item \py{action_over_T}: the value of the action divided by the temperature, i.e. $S_{E,d} / T$;
    \item \py{phi0_min}: the field value $\phi_{0}$ minimising $S_{E,d}$ in~\Cref{eq:action};
    \item  \py{V_exit}: the value of the potential at $V(\phi_{0})$;
    \item \py{true_min}: the value of the field $\phi$ at the true vacuum;
    \item \py{phi_original_false_vev}: the value of the field $\phi$ at the false vacuum;
    \item \py{min_V}: the value of the potential at the true vacuum.
\end{itemize}
The boolean attribute \py{barrier} indicates whether a barrier is present at the specified temperature.

The class \py{espinosa.Vt_vec} accepts the optional arguments \py{d} (set to \py{3} by default) for the number of dimensions, \py{vev0} (default \py{100}, assumed to be in the same energy units as the field entering in the potential) for an initial estimate of the finite temperature vev value, \py{step_phi} (default \py{1e-3}) that sets the step size in the $\phi$ dimension in units of \py{vev0}, \py{precision} (default \py{1e-3}) determining the desired relative precision in the computation of $S_{E,d}$, \py{ratio_vev_step0} (default \py{50}) that sets the initial step size in the $\widetilde{\phi}_0$ dimension as \py{vev0 / ratio_vev_step0}. In addition, the optional argument \py{save_all} (default \py{False}) determines whether the internal computations for each value of $\widetilde{\phi}_0$ should be saved as an attribute of the class instance; this can be useful if the user wants to perform consistency checks on the results or post-process these data.

\begin{figure}[htb]
    \centering
    \includegraphics[width=0.99\linewidth]{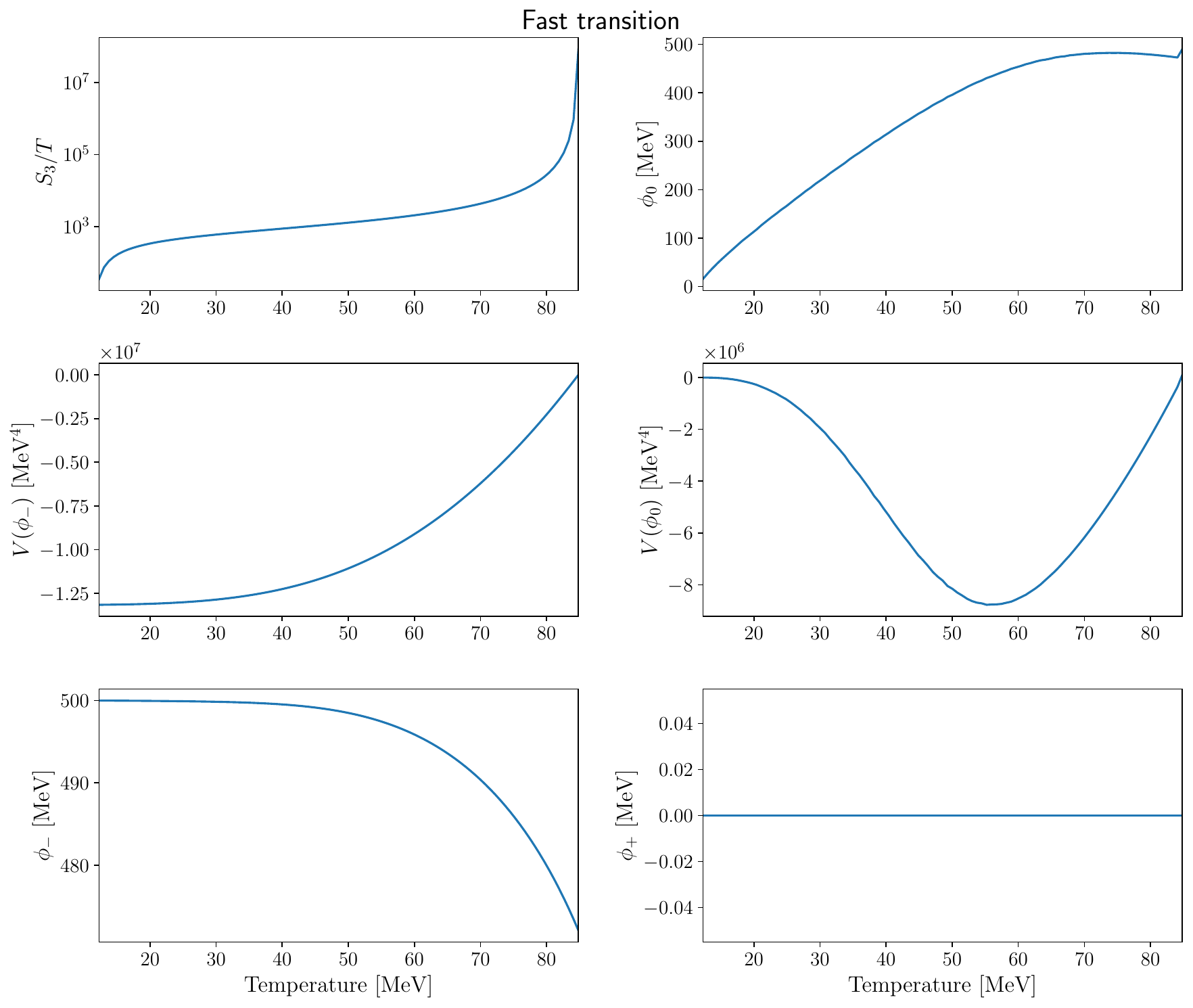}
    \caption{
    Evolution of the relevant quantities related to tunnelling as a function of temperature for the ``Fast'' point in~\Cref{tab:input}. 
    \emph{Upper-left}: action $S_{E,3} / T$. 
    \emph{Upper-right}: tunnelling solution $\phi_0$. 
    \emph{Centre-left}: potential at true vacuum $V(\phi_-)$.
    \emph{Centre-right}: potential at $\phi_0$, $V(\phi_0)$.
    \emph{Bottom-left}: true vacuum $\phi_-$.
    \emph{Bottom-right}: false vacuum $\phi_+$.
    }
    \label{fig:action_fast}
\end{figure}

\begin{figure}[htb]
    \centering
    \includegraphics[width=0.99\linewidth]{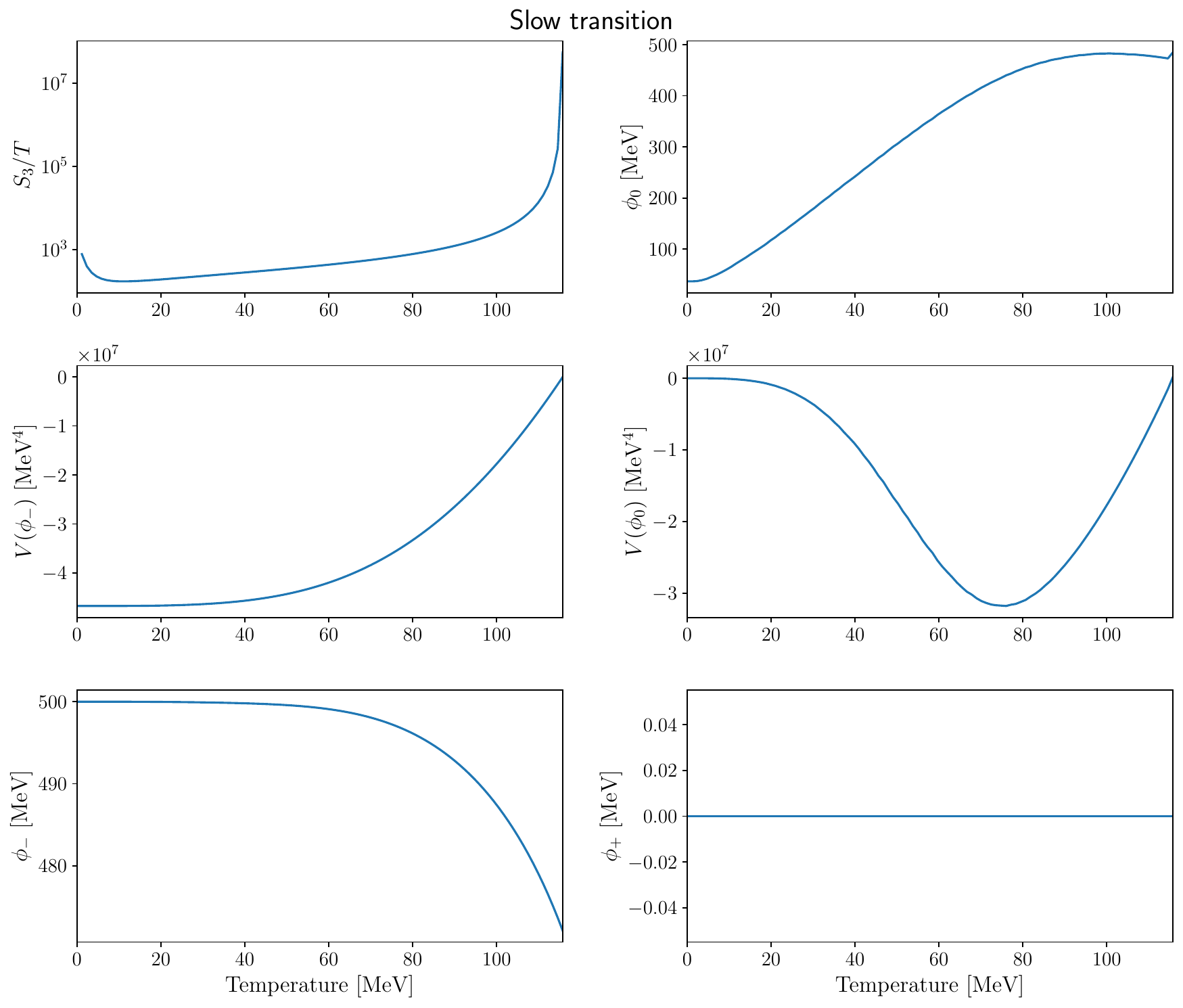}
    \caption{Same as in~\Cref{fig:action_fast}, but for the ``Slow'' point in~\Cref{tab:input}.}
    \label{fig:action_slow}
\end{figure}

After having defined the function \py{action_over_T()}, the rest of the code in~\Cref{code:action_over_T} simply computes the tunnelling quantities over 100 values of temperature between $T_{min}$ and $T_c$. This computation, performed on an Apple M2 processor, employs approximately 2 seconds, thus amounting to approximately 20 milliseconds for each tunnelling computation. The results of the computation are reported in~\Cref{fig:action_fast} for the ``Fast'' transition benchmark point, and in~\Cref{fig:action_slow} for the ``Slow'' one. Each plot shows, as a function of the temperature $T$, the action $S_{E,3} / T$ (upper-left), the solution $\phi_0 (T)$ (upper-right)\footnote{We remind the reader that $\phi_0$ is the value of the field at the centre of the bubble right after it is nucleated.} and the potential value at this point $V(\phi_0(T), T)$ (middle-right), the position of the true vacuum $\phi_-(T)$ (bottom-left) and the minimum of the potential $V(\phi_-(T), T)$ (middle-left panel), as well as the false vacuum value $\phi_+(T)$ (bottom-right panel). The most striking difference between the two points is in the shape of the action. It monotonically decreases with temperature for the fast transition, while in the slow scenario, it shows a global minimum at $T=11.7$ MeV. 

\subsection{Computation of the fraction of Universe in the true and false vacua}

The key quantity that tracks the evolution of the transition at each temperature value is the fraction of the Universe in the false vacuum $P_f(T)$ (or equivalently the fraction of volume in true vacuum $P_t(T) = 1 - P_f(T)$). Bubble nucleation is a stochastic process that happens continuously as soon as it is energetically allowed, with a nucleation probability per unit volume and unit time given by~\Cref{eq:nuc_rate}. Although for fast transitions there is usually a characteristic temperature that fully defines the dynamics of the process (the so-called nucleation temperature, where most of the bubble nucleation happens), this is generally not true for slow transitions, which are phenomenologically more interesting since they produce GW signals with larger amplitudes and therefore better detection prospects. 

To obtain general, reliable results, one shall thus track the evolution of $P_f(T)$ over the full range of temperatures where bubble nucleation is possible, given by $T\in[T_{min}, T_c]$. The efficient computation of the bounce action in \elena makes it possible to study the phase transition following the evolution of $P_f(T)$, instead of relying on approximations such as comparing the nucleation rate in~\Cref{eq:nuc_rate} to the Hubble expansion rate, $H(T)$.

\elena provides modules to readily compute $P_f(T)$ from the tunnelling action, without making any strong assumption on the underlying cosmology. The only underlying assumptions are the ones leading to the JMAK equation~\cite{kolmogorov1937statistical,johnson1939reaction,avrami1939kinetics,avrami1940kinetics,avrami1941granulation,Guth:1979bh,Guth:1981uk}, which is given by:
\begin{equation}\label{eq:JMAK}
    P_f(T) = \exp \left(- \mathcal{V}_t^{ext} (T) \right),
\end{equation}
where $\mathcal{V}_t^{ext} (T)$ is the \emph{extended} volume in true vacuum at a given temperature. The extended volume is defined as the volume that all nucleated bubbles would occupy, assuming a nucleation probability given by $\Gamma(T)$, while neglecting the fact that a fraction of this volume may have already been converted to true vacuum at that temperature. 
This amounts to considering processes such as the nucleation of a new bubble inside a region of true vacuum, or double-counting the volume when two bubbles grow into each other. $\mathcal{V}_t^{ext} (T)$ clearly gives a strong overestimate of the real fraction of volume in true vacuum (from which the ``extended'' adjective). 
Nevertheless, it has been demonstrated that, under reasonable assumptions, the fraction of the Universe in the false vacuum state is correctly related to $\mathcal{V}_t^{ext}$ via~\Cref{eq:JMAK}. We refer the reader to Ref.~\cite{Athron:2023xlk} for a thorough discussion on the JMAK equation in the context of cosmological phase transitions, including multiple derivations of it and a critical analysis of the validity of its assumptions.  

The extended volume in true vacuum is obtained by summing the volumes of the single bubbles nucleated up to a given time $t$. By assuming a homogeneous rate of bubble nucleation per unit volume and unit time $\Gamma(t)$, the corresponding expression in a static Universe reads
\begin{equation}
    \mathcal{V}_{t,\mathrm{static}}^{ext} = \int_{t_0}^t \de t' \Gamma\left(t'\right) V \left(t', t\right),
\end{equation}
where $t_0$ is the time corresponding to the critical temperature $T_c$ and $V(t',t)$ is the volume at time $t$ of a bubble that nucleated at time $t'$. Under the assumption of spherically symmetric bubbles, it is given by
\begin{equation}
V(t,t') = \frac{4\pi}{3}  R(t,t')^3,
\end{equation}
with the radius at time $t$ given by
\begin{equation}
R(t,t') = R_{0}(t') +  \int_{t'}^t dt'' v_{w}(t''),
\end{equation}
where $R_0$ is the initial bubble radius and $v_w$ the bubble wall velocity. Assuming that a bubble rapidly grows to a radius much larger than $R_{0}(t')$, and that the wall rapidly accelerates to its final constant velocity $v_{w}$, taken to be a homogeneous parameter in the Universe, one gets
\begin{equation}
V(t,t') = \frac{4\pi}{3} \left[ v_{w} \int_{t'}^t dt'' \right]^3,
\end{equation}
where we keep the explicit form of the time integral for subsequent generalisation to the case of an expanding Universe. The expression for the extended volume thus reads
\begin{equation}
\mathcal{V}^{ext}_{t,\mathrm{static}}(t) = \frac{4\pi}{3} v_{w}^3 \int_{t_{0}}^t dt' \Gamma(t') \left[ \int_{t'}^{t} dt''  \right]^3.
\end{equation}

The generalisation to an expanding universe is readily obtained. First, we must appropriately scale the nucleation rate to reflect the change in unit volume over time
\begin{equation}
\int_{t_{0}}^t dt' \Gamma(t') \rightarrow \int_{t_{0}}^t dt' \Gamma(t') \left( \frac{a(t')}{a(t)} \right)^3,
\end{equation}
where we have chosen to normalise the unit volume at time $t$. 
Then, we must account for the fact that the radius of a bubble increases both because of proper motion and because of the expansion of the Universe
\begin{equation}
\int_{t'}^t dt'' \rightarrow \int_{t'}^t dt'' \frac{a(t)}{a(t'')}.
\end{equation}
Thus, the extended volume in an expanding universe finally reads
\begin{equation}\label{eq:v_ext_time}
\mathcal{V}^{ext}_{t}(t) = \frac{4\pi}{3} v_{w}^3 \int_{t_{0}}^t dt' \Gamma(t') \left( \frac{a(t')}{a(t)} \right)^3 \left[ \int_{t'}^{t} dt'' \frac{a(t)}{a(t'')}  \right]^3.
\end{equation}

One would like to express~\Cref{eq:v_ext_time} as a function of temperature, given that dynamical quantities relevant to phase transition processes in the early Universe are more conveniently computed as a function of temperature rather than time. A common assumption usually employed in the literature is the use of the MIT bag equation of state~\cite{Chodos:1974je}, under which the following equalities hold
\begin{equation}
    \frac{dt}{dT} \overset{\scalebox{0.5}{\text{bag}}}{=} - \frac{1}{H(T) T},\hspace{1cm} \frac{a(t_{1})}{a(t_{2})} \overset{\scalebox{0.5}{\text{bag}}}{=} \frac{T_{2}}{T_{1}}.
\end{equation}
Moreover, the energy density of the Universe is often taken to be radiation dominated, setting the functional form for the Hubble expansion rate $H(T)$. These assumptions are valid for fast transitions in which the amount of vacuum energy released in the FOPT is subdominant with respect to the radiation energy density at the time. However, they do not necessarily hold for strong supercooled phase transitions, in which most of the Universe remains in the false vacuum during a period in which the temperature can drop significantly. Given that the radiation contribution scales as $\rho \propto T^4$ while the vacuum energy density $V(\phi_+, T) - V(\phi_-, T)$ generally increases for smaller temperatures (cf. e.g.~\Cref{fig:action_fast,fig:action_slow}),
strong supercooled phase transitions can exhibit periods of vacuum domination, or at least feature regimes where the vacuum and radiation contributions are comparable. 

To properly account for these scenarios, \elena employs a more general relation~\cite{Athron:2022mmm} between time and temperature that only assumes the adiabatic expansion of the Universe, and includes the contributions from both the Standard Model (SM) and BSM fields to the potential, in order to evolve the temperature over time
\begin{equation}\label{eq:general_T_t}
\frac{dT}{dt} = - 3 H(T) \frac{\partial_{T} V^\textrm{T}(\phi_{+} (T), T)}{\partial_{TT} V^\textrm{T}(\phi_{+} (T), T)}, \hspace{1cm} \frac{a(t_{1})}{a(t_{2})} = \exp \left( \int_{t_{2}}^{t_{1}} dt' H(t') \right),
\end{equation}
where $\partial_T V = \partial V / \partial T$, $\partial_{TT} V = \partial^2 V / \partial T^2$ and $V^\textrm{T}$ is meant to refer to the total potential, including the ($T$-dependent but $\phi$-independent) SM component and the contribution from BSM fields.
The expression used in \elena for the computation of the extended volume thus reads
\begin{equation}
\mathcal{V}^{ext}_{t}(T) = \frac{4\pi}{3} v_{w}^3 \int_{T}^{T_{c}} dT' \frac{\Gamma(T')}{3 H(T')} \frac{\partial_{TT} V^\textrm{T}(T')}{\partial_{T} V^\textrm{T}(T')} \left( \frac{a(T')}{a(T)} \right)^3 \left[ \int_{T}^{T'} dT'' \frac{1}{3 H(T'')} \frac{\partial_{TT} V^\textrm{T}(T'')}{\partial_{T} V^\textrm{T}(T'')} \frac{a(T)}{a(T'')}  \right]^3,
\end{equation}
with the ratio of scale factors given by
\begin{equation}\label{eq:a_ratio_general}
\frac{a(T_{1})}{a(T_{2})} = \exp \left( \int_{T_{1}}^{T_{2}} dT' \frac{1}{3} \frac{\partial_{TT} V^\textrm{T}(T')}{\partial_{T} V^\textrm{T}(T')} \right).
\end{equation}

The actual computation is performed by the function \py{temperatures.compute_logP_f}, as exemplified in the~\Cref{code:log_Pf}.
\begin{lstlisting}[language=Python, caption=Computation of the fraction of Universe volume in false vacuum., label=code:log_Pf]
from temperatures import compute_logP_f

logP_f, Temps, ratio_V, Gamma, H = compute_logP_f(dp, V_min_value, S3overT, v_w = 1, units = 'MeV')
\end{lstlisting}
The function \py{temperatures.compute_logP_f} takes as required arguments the model class instance and two dictionaries, containing for each key temperature value the corresponding value of the potential at the true minimum and the tunnelling action (\py{dp, V_min_value, S3overT} in the~\Cref{code:log_Pf}), computed in the~\Cref{code:model,code:action_over_T} in our example. Optional arguments are the asymptotic wall velocity (\py{v_w}, set by default to \py{1}) and the units for dimensionful quantities (\py{units}, default \py{'GeV'}).
The function returns the natural logarithm of $P_f(T)$ as an array (\py{logP_f}), where each entry corresponds to its value at the temperature reported in the same entry in the array \py{Temps} (sorted in increasing values of temperature). For convenience, the function also returns the arrays \py{ratio_V, Gamma, H}, which contain the values of $\partial_{TT} V^\textrm{T} / \partial_{T} V^\textrm{T} (T)$, $\Gamma(T)$ and $H(T)$ for each temperature contained in \py{Temps}, so that they can be stored and used in other computations or consistency checks.

\subsection{Computation of transition milestone temperatures}
The knowledge of $P_f(T)$ allows for the computation of the temperatures at which important milestones in the transition are reached. We show in this section how \elena can be used to compute the \textbf{nucleation}, \textbf{percolation} and \textbf{completion} temperatures, taking into account the full evolution of the transition.

The nucleation temperature $T_n$ is defined as the moment when there is, on average, one nucleated bubble per Hubble volume. A commonly adopted heuristic criterion for estimating $T_n$ in the literature is
\begin{equation}
    \frac{\Gamma\left(T_n\right)}{H^4\left(T_n\right)} \equiv 1,
\end{equation}
where $\Gamma$ is the false vacuum decay rate~\Cref{eq:nuc_rate} and $H$ is the Hubble parameter. For transitions at the electroweak scale, an even simpler estimation is often adopted~\footnote{This is for example the default criterion in \CT, see e.g.~\cite{Wainwright:2011kj}.}, given by the condition $S_{E,3}(T_N) / T_N \sim 140$.

Despite being a commonly employed quantity to characterise the temperature at which a FOPT takes place, the nucleation temperature does not represent a relevant physical quantity in the characterisation of the FOPT dynamics; its extended usage in the literature has mainly been due to its simple numerical implementation, given that it does not require the determination of the false vacuum fraction. Indeed, it is possible to find realisations in which the phase transition takes place but the nucleation criterion is never met~\cite{Athron:2022mmm}, particularly in the context of supercooled phase transitions. Nevertheless, we stress here that the nucleation temperature gives the correct estimate for the relevant transition temperature in the case of fast transitions.

\elena provides the function \py{temperatures.N_bubblesH} to compute the average number of nucleated bubbles per Hubble volume as a function of temperature, by using the general equation~\cite{Athron:2023xlk}
\begin{equation}\label{eq:nucleation_general}
    N(t) = \frac{4 \pi}{3} \int_{t_c}^t \de t' \frac{\Gamma\left( T\left(t'\right) \right) P_f\left(t'\right)}{H^3 \left(t'\right)},
\end{equation}
from which the nucleation time $t_n$ can be derived by looking for the solution $N(t_n) = 1$.
The usage of \py{temperatures.N_bubblesH} is exemplified in the~\Cref{code:n_bubbles}.
\begin{lstlisting}[language=Python, caption=Computation of the average number of nucleated bubbles per Hubble volume., label=code:n_bubbles]
from temperatures import compute_logP_f, N_bubblesH

nH = N_bubblesH(Temps, Gamma, logP_f, H, ratio_V)
\end{lstlisting}
The function \py{temperatures.N_bubblesH} takes as arguments a series of arrays \py{Temps, Gamma, logP_f, H, ratio_V}, corresponding to the output of the function \py{temperatures.compute_logP_f}, and returns an array containing the average number of bubbles per Hubble volume at each temperature value in \py{Temps}.

Percolation is defined as the moment when a connected cluster of bubbles that spans the entire Universe exists. The temperature at which it takes place characterises the peak frequency in the gravitational wave spectrum from a FOPT~\cite{Athron:2023xlk}, given that at percolation the majority of bubbles responsible for the GW production have collided. It has been shown that a fully connected cluster of bubbles almost certainly forms when a specific value of the true vacuum fraction has been reached~\cite{Broadbent:1957rm,Shante01051971,hunt2014percolation}; for cosmological phase transitions featuring uniformly nucleated spherical bubbles, this threshold value is $P_t \approx 0.29$~\cite{10.1063/1.1338506,LIN2018299,LI2020112815}. Therefore, the percolation temperature $T_p$ is defined by the equation
\begin{equation}
    P_f\left( T_p \right) = 0.71.
\end{equation}
Percolation studies performed in the framework of solid state physics assume a static background, but in cosmological scenarios, the expansion of the Universe must be taken into account, especially in the case of slow supercooled phase transitions. In an expanding Universe, the expansion of bubbles towards each other competes with the expansion of space itself, which tends to increase the physical distance among bubbles. If the latter dominates, the bubbles might actually not collide, even if the fraction of the Universe in the false vacuum $P_f$ decreases with time. Therefore, a necessary condition for the production of GW is that the physical volume in false vacuum, given by~\cite{Turner:1992tz} 
\begin{equation}
    \mathcal{V}_\mathrm{phys} \left( t \right) = a^3 \left( t \right) P_f \left( t \right),
\end{equation}
is decreasing when the bubbles are assumed to collide, which corresponds to
\begin{equation}\label{eq:dVphys_dt}
    \frac{\de \mathcal{V}_\mathrm{phys}}{\de t} = \mathcal{V}_\mathrm{phys} \left( t \right) \left [\frac{\de}{\de t} \ln \left(P_f \left(t \right) \right) + 3 H \left( t \right)  \right] \leq 0.
\end{equation}
We can track the evolution of $\mathcal{V}_\mathrm{phys}(T)$ with temperature by substituting~\Cref{eq:general_T_t} into~\Cref{eq:dVphys_dt}, from which one finds that the physical volume in the false vacuum is decreasing at a given temperature if the condition
\begin{equation}\label{eq:volume_decreasing}
    \frac{\de \ln P_f}{\de T} \geq \frac{\partial_{TT} V^\textrm{T}}{\partial_T V^\textrm{T}}
\end{equation}
is satisfied.

The percolation of bubbles in an expanding universe is questionable if $\mathcal{V}_\mathrm{phys}$ is not decreasing at $T_p$~\cite{Turner:1992tz,Athron:2022mmm,Ellis:2018mja}, although it is worth noticing that percolation studies in an expanding spacetime have not been performed to this date~\cite{Athron:2023xlk}, leaving open the question on how significantly the results can change compared to the static scenario.

Finally, the completion temperature is defined as the one at which the value of $P_f$ has decreased below a certain threshold value, signalling that most of the Universe has been converted to the true vacuum~\footnote{We use the subscript $e$ for ``end'', to avoid confusion with the commonly used $T_c$ notation that refers to the critical temperature.}. It is given by
\begin{equation}\label{eq:condition_physical_volume}
    P_f\left( T_{e} \right) = \epsilon,
\end{equation}
with $\epsilon \ll 1$. The value of $T_e$ generally depends weakly on the choice of $\epsilon$, except when the asymptotic limit of $P_f$ is different from zero. A common choice is to require that at least $99\%$ of the Universe has been converted to the true vacuum, $\epsilon = 0.01$.

The full computation of the transition milestones and verification of decreasing physical volume in the false vacuum at percolation in \elena is exemplified in the~\Cref{code:milestones}. 
\begin{lstlisting}[language=Python, caption=Computation of the transition temperatures milestones and of the evolution of physical volume in false vacuum., label=code:milestones]
from utils import interpolation_narrow
from temperatures import compute_logP_f, N_bubblesH

is_physical = True

def is_increasing(arr):
    return np.all(arr[:-1] <= arr[1:])

counter = 0
while counter <= 1:
    if counter == 1:
        temperatures = np.linspace(np.nanmax([T_min, 0.95 * T_completion]), np.nanmin([T_max, 1.05 * T_nuc]), n_points, endpoint = True)
        action_vec(temperatures)
    logP_f, Temps, ratio_V, Gamma, H = compute_logP_f(dp, V_min_value, S3overT, v_w = 1, units = units)
    nH = N_bubblesH(Temps, Gamma, logP_f, H, ratio_V)
    mask_nH = ~np.isnan(nH)
    T_nuc = interpolation_narrow(np.log(nH[mask_nH]), Temps[mask_nH], 0)
    mask_Pf = ~np.isnan(logP_f)
    T_perc = interpolation_narrow(logP_f[mask_Pf], Temps[mask_Pf], np.log(0.71))
    T_completion = interpolation_narrow(logP_f[mask_Pf], Temps[mask_Pf], np.log(0.01))
    
    idx_compl = np.max([np.argmin(np.abs(Temps - T_completion)), 1])
    test_completion = np.array([logP_f[idx_compl - 1], logP_f[idx_compl], logP_f[idx_compl + 1]])
    test_completion = test_completion[~np.isnan(test_completion)]

    if not is_increasing(test_completion):
        T_completion = np.nan
    if counter == 1:
        d_dT_logP_f = np.gradient(logP_f, Temps)
        log_at_T_perc = interpolation_narrow(Temps, d_dT_logP_f, T_perc)
        ratio_V_at_T_perc = interpolation_narrow(Temps, ratio_V, T_perc)
        log_at_T_completion = interpolation_narrow(Temps, d_dT_logP_f, T_completion)
        ratio_V_at_T_completion = interpolation_narrow(Temps, ratio_V, T_completion)
        if ratio_V_at_T_perc > log_at_T_perc:
            is_physical = False
    counter += 1
\end{lstlisting}
The code in~\Cref{code:milestones} defines an auxiliary boolean flag \py{is_physical}, which will be set to \py{True} if the final SGWB prediction is reliable, and \py{False} if some underlying assumption in the calculation is violated (more details on these conditions are presented in the following). The function \py{is_increasing} simply verifies if a \textsf{NumPy} array is monotonically increasing, and is used as a further check on the completion of the transition.
The \py{while} loop is implemented to improve the precision of the computation: the milestone temperatures are first calculated by employing the values of the action computed over the full range of temperatures between $T_{min}$ and $T_c$; once a first estimate of $T_e$ and $T_n$ is obtained, the action is computed over the temperature range $T \in [0.95\ T_e, 1.05\ T_n]$ with a smaller step size in temperature, to increase the precision of the integrations appearing in~\Cref{eq:general_T_t,eq:a_ratio_general,eq:nucleation_general} over the most relevant temperature interval. Notice that this is an optional improvement: the user can skip this further computation and simply use the output from the first iteration, depending on the desired trade-off between computational time and precision.
The rest of the code uses the already discussed functions \py{temperatures.compute_logP_f} and \py{temperatures.N_bubblesH} to compute $\ln(P_f(T))$ (\py{logP_f}) and $N(T)$ (\py{nH}), and some \textsf{NumPy} masks are defined to select only temperatures where the result of the computation is a float number (\py{mask_Pf} and \py{mask_nH}) to prevent potential issues in the subsequent computations due to the presence of \textsf{NumPy} \texttt{Not a Number (NaN)} (if any).
The function \py{utils.interpolation_narrow} is a wrapper of \py{numpy.interp} that restricts the range of interpolation to avoid numerical problems that can result from the very large range of values that the functions $\ln(P_f(T))$ and $N(T)$ can span. It is called as \py{utils.interpolation_narrow(y, x, target)}, where \py{y} and \py{x} are \textsf{NumPy} arrays of the same dimension, and \py{y} is interpreted as a function of \py{x}; the function returns the value of the independent variable \py{x} for which the variable \py{y} is equal to \py{target}. If no such solution is found, it returns a \py{numpy.nan}. The code uses \py{utils.interpolation_narrow} to compute the nucleation (\py{T_nuc}), percolation (\py{T_perc}) and completion (\py{T_completion}) temperatures.
The following lines isolate the range of $\ln(P_f(T))$ values immediately around the completion temperature (\py{test_completion}) and check if the value of $\ln(P_f(T))$ is decreasing at that temperature, to ensure that the fraction of volume in false vacuum is evolving towards smaller values when $P_f = \epsilon$. If this condition is not satisfied, the variable \py{T_completion} is set to \py{numpy.nan}, to indicate that a valid completion criterion was not found.
Finally, the code computes $\de \ln P_f / \de T$ (\py{d_dT_logP_f}) over the full range of temperatures, as well as the values of $\de \ln P_f / \de T$ and $\partial_{TT} V^\textrm{T} / \partial_T V^\textrm{T}$ at percolation (\py{log_at_T_perc} and \py{ratio_V_at_T_perc}) and completion (\py{log_at_T_completion} and \py{ratio_V_at_T_completion}). In the case in which the condition in~\Cref{eq:volume_decreasing} is violated, the flag \py{is_physical} is set to \py{False}.

\begin{figure}[htb]
    \centering
    \includegraphics[width=0.99\linewidth]{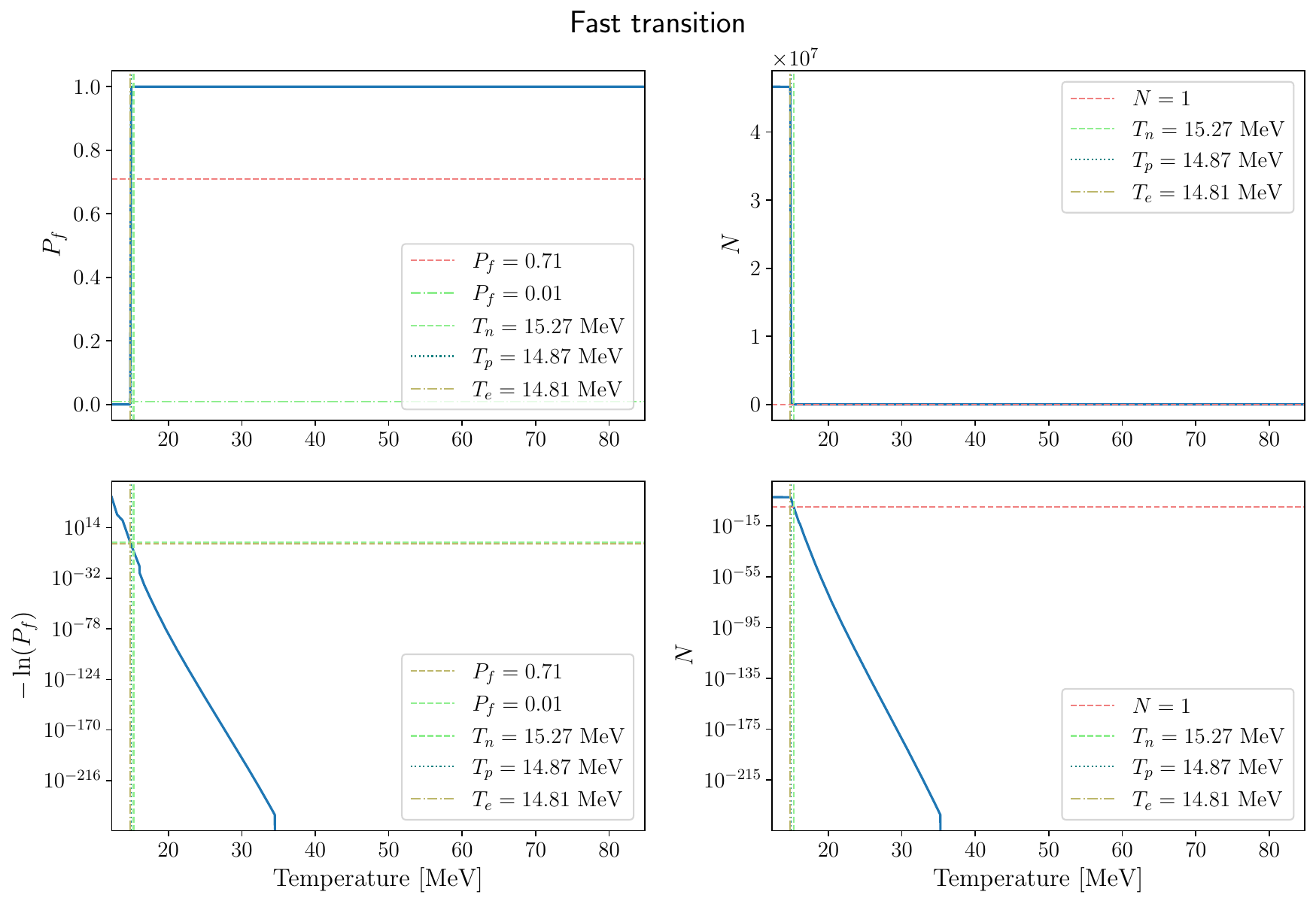}
    \caption{Evolution of the transition for the ``Fast'' benchmark point in~\Cref{tab:input}. \emph{Left panels} show the value of $P_f(T)$ (top) and $-\ln P_f(T)$ (bottom). \emph{Right panels} show the evolution of $N(T)$ on linear (top) and logarithmic (bottom) scales. The vertical lines signal the values of the nucleation ($T_n$, dashed line), percolation ($T_p$, dotted line) and completion ($T_{e}$, dot-dashed line) temperatures. The horizontal lines on the left panels signal the values of $P_f = 0.71$ (dashed line) and $P_f = 0.01$ (dot-dashed line), while on the right panel they signal the value $N = 1$. 
    }
    \label{fig:evolution_fast}
\end{figure}

We report in~\Cref{fig:evolution_fast} the results of the computation for the ``Fast'' benchmark in~\Cref{tab:input}. The left panels show the value of the fraction of the Universe in false vacuum $P_f$ (top) and of $-\ln P_f$ (bottom). Vertical lines signal the values of the nucleation ($T_n$, dashed line), percolation ($T_p$, dotted line) and completion ($T_{e}$, dot-dashed line) temperatures, while horizontal lines signal the values of $P_f = 0.71$ (dashed line) and $P_f = 0.01$ (dot-dashed line). On the right panels we report the evolution of the average number of bubbles per Hubble volume $N$, both on linear scale (top panel) and logarithmic scale (lower panel).
Vertical lines follow the same convention as on the left panels, while the horizontal dashed line corresponds to the value $N = 1$.
\begin{figure}[htb]
    \centering
    \includegraphics[width=0.99\linewidth]{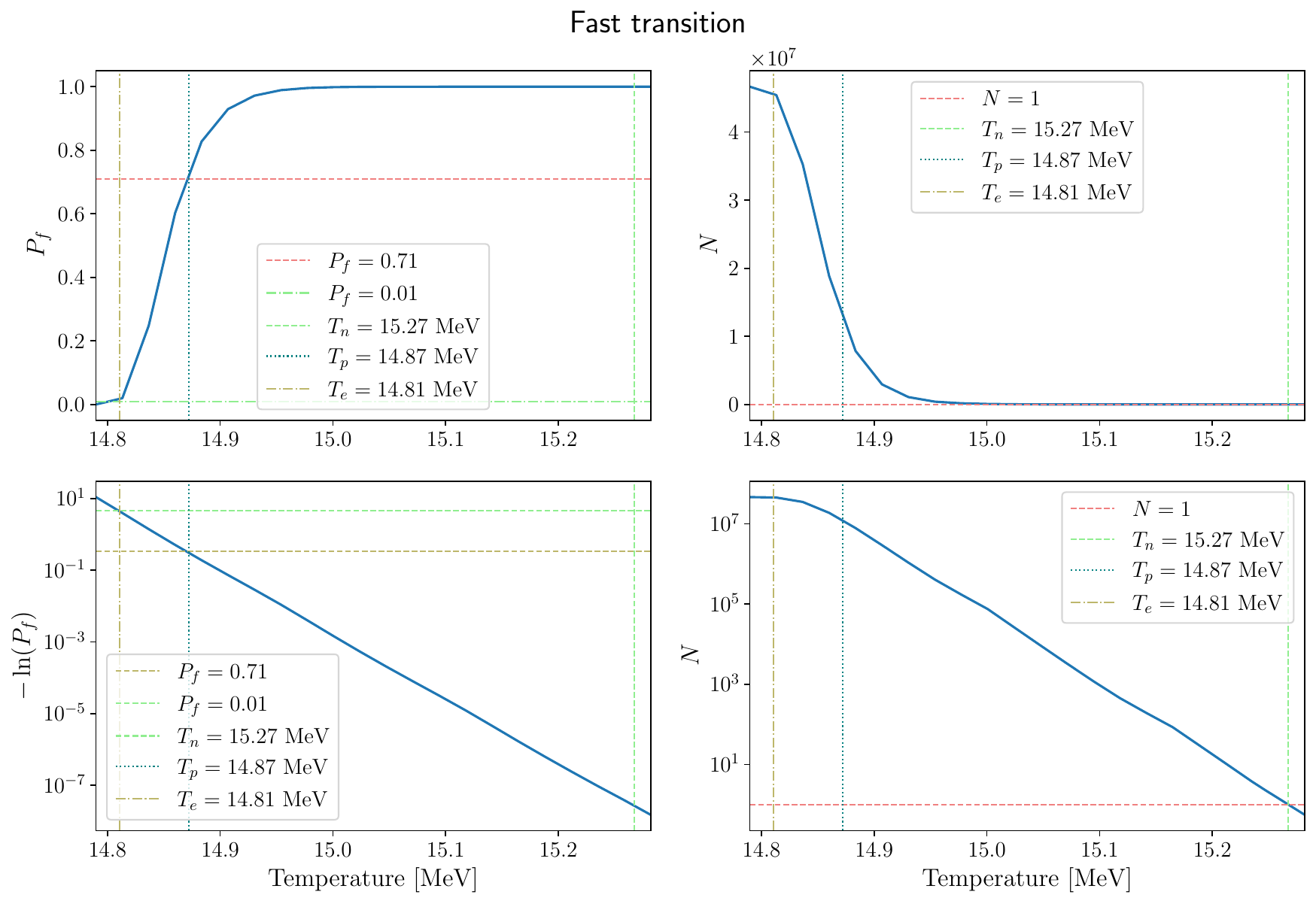}
    \caption{Same as~\Cref{fig:evolution_fast}, but zoomed over the range of temperatures between nucleation and completion.}
    \label{fig:evolution_fast_zoom}
\end{figure}
\Cref{fig:evolution_fast_zoom} shows the same quantities as in~\Cref{fig:evolution_fast}, but zoomed over the most interesting range of temperatures between nucleation and completion.

\begin{figure}[htb]
    \centering
    \includegraphics[width=0.99\linewidth]{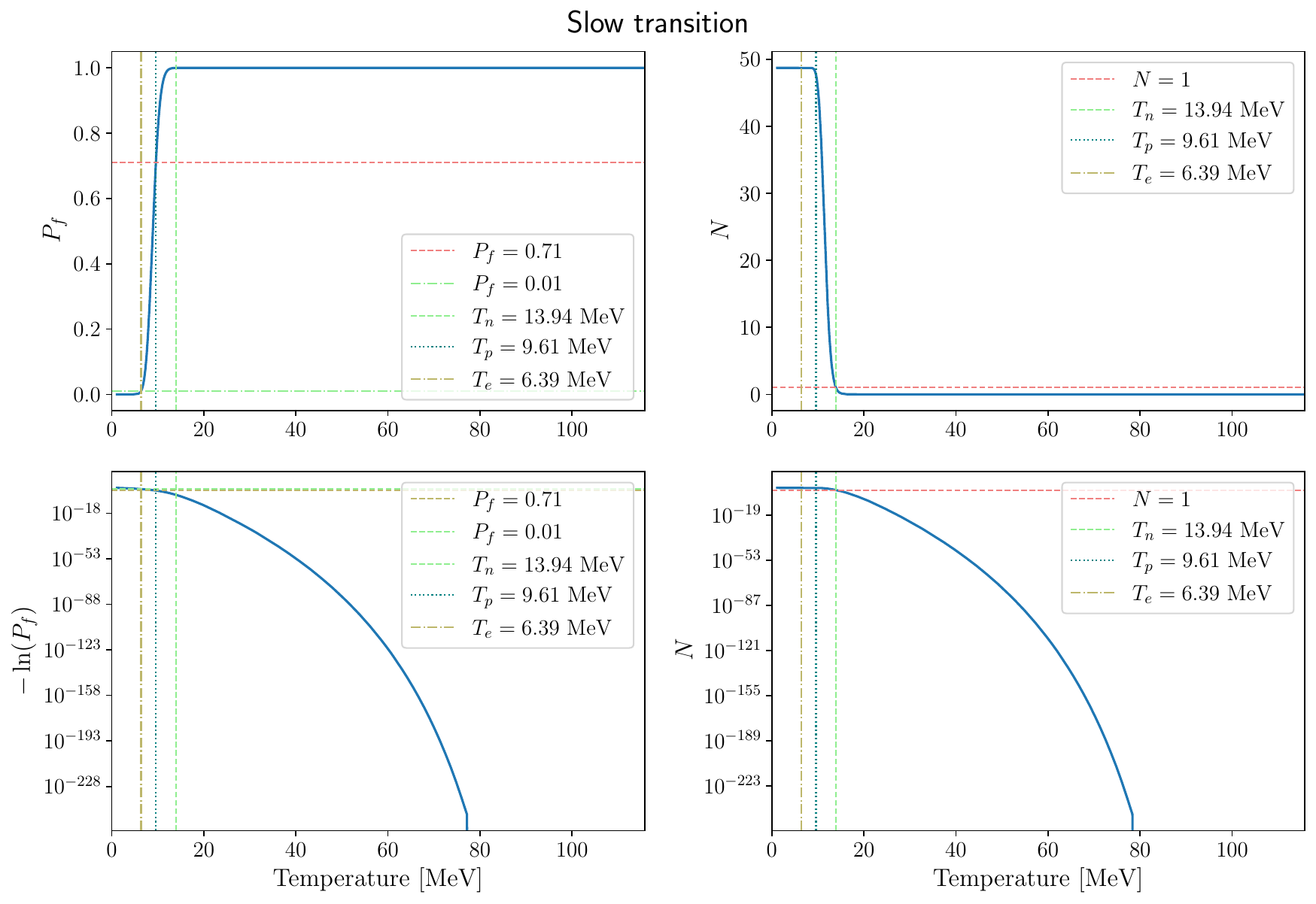}
    \caption{Same as in~\Cref{fig:evolution_fast}, but for the ``Slow'' benchmark in~\Cref{tab:input}.}
    \label{fig:evolution_slow}
\end{figure}

\begin{figure}[htb]
    \centering
    \includegraphics[width=0.99\linewidth]{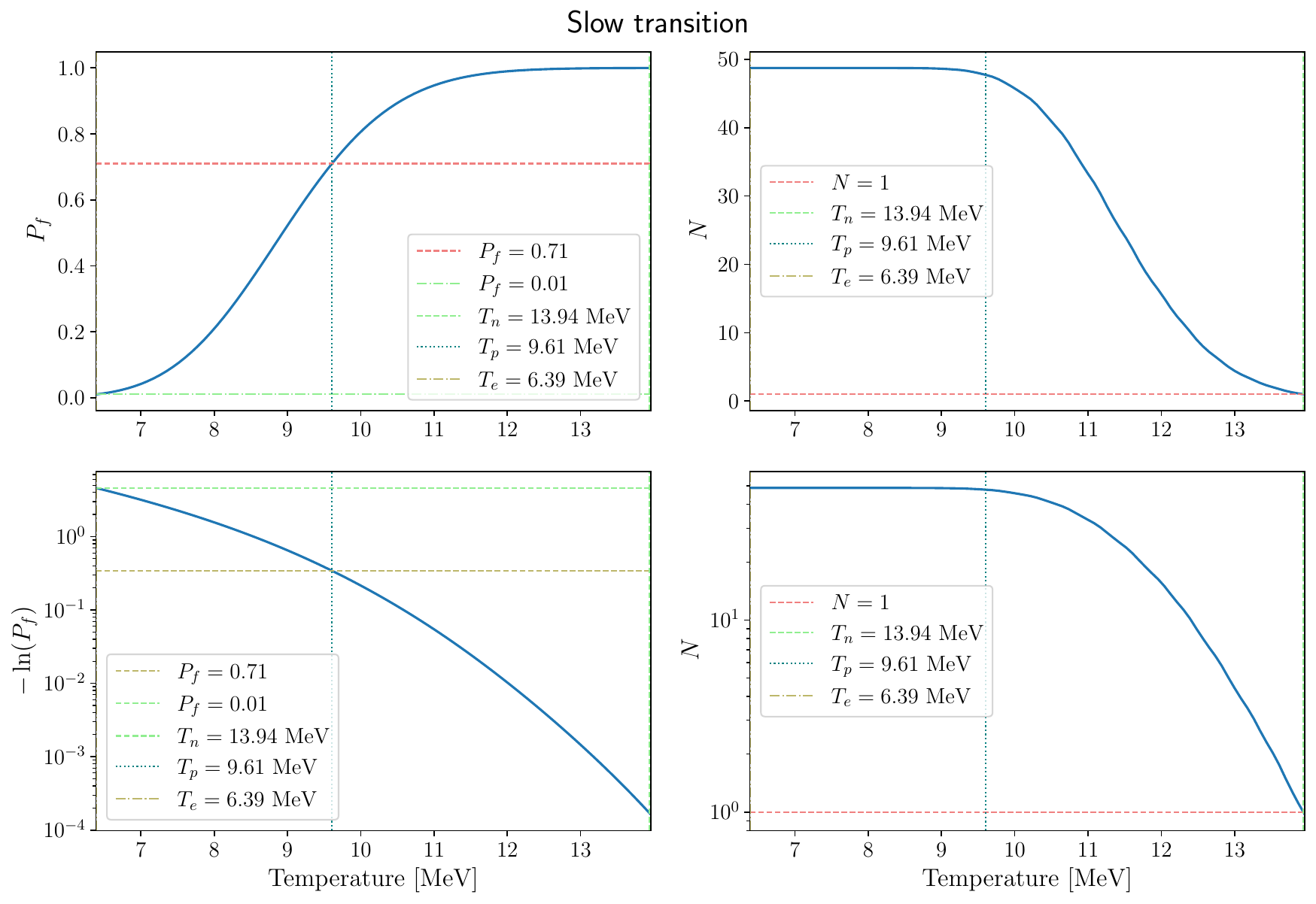}
    \caption{Same as~\Cref{fig:evolution_slow}, but zoomed over the range of temperatures between nucleation and completion.}
    \label{fig:evolution_slow_zoom}
\end{figure}

The same quantities are plotted in~\Cref{fig:evolution_slow,fig:evolution_slow_zoom} for the ``Slow'' benchmark in~\Cref{tab:input}.

The most striking difference between the two benchmark points in~\Cref{tab:input} is obviously the duration of the transition: in the ``Fast'' transition, the nucleation of bubbles becomes sizeable at $T_n^\textrm{fast} = 15.27$ MeV, and the full transition completes shortly after, at $T_e^\textrm{fast} = 14.81$ MeV. Contrary, in the ``Slow'' transition the nucleation is sizeable at $T_n^\textrm{slow} = 13.94$ MeV, and the transition completes at $T_e^\textrm{slow} = 6.39$ MeV. Notice also the difference in the average number of bubbles per Hubble volume $N$, which reaches an asymptotic value of $N(T_{min})^\textrm{fast} = 4.67 \times 10^7$ in the ``Fast'' benchmark, and $N(T_{min})^\textrm{slow} = 48.73$ in the ``Slow'' one. This generally translates into larger SGWB signals from slow transitions, given that the colliding bubbles are larger and thus contain a larger amount of energy to produce GW.

\begin{figure}[htb]
    \centering
    \includegraphics[width=0.99\linewidth]{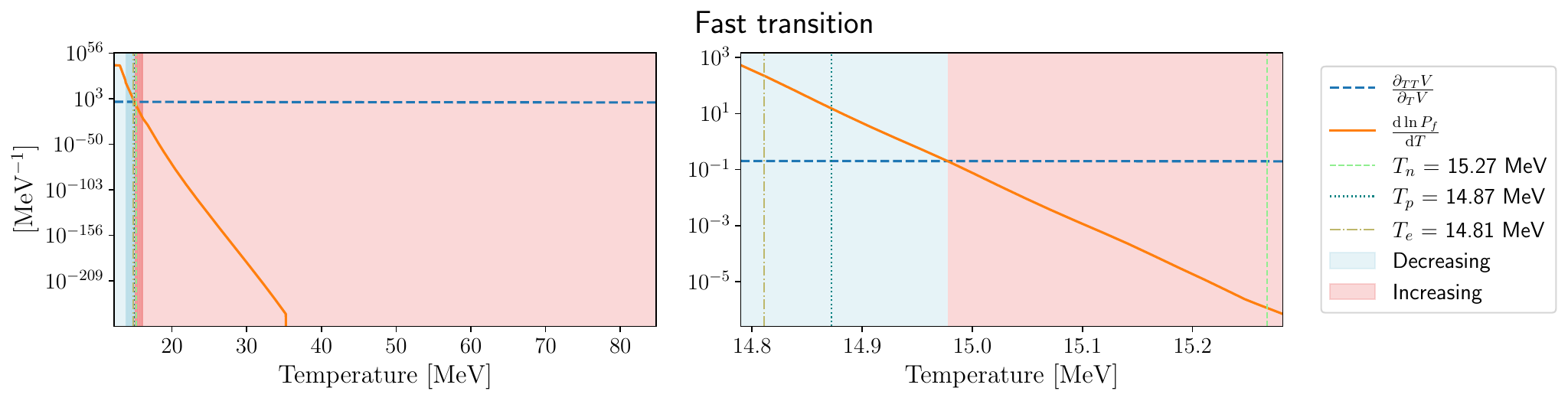}
    \caption{The evolution of $\de \ln P_f / \de T$ (continuous curve) and $\partial_{TT} V^\textrm{T} / \partial_T V^\textrm{T}$ (dashed curve) with temperature for the ``Fast'' benchmark in~\Cref{tab:input}. Temperature values where the condition in~\Cref{eq:volume_decreasing} is satisfied (violated) are highlighted by a light-blue (light-coral) background. The vertical dashed, dotted and dot-dashed lines represent the values of the nucleation  ($T_n$), percolation ($T_p$) and completion ($T_e$) temperatures, respectively. The left panel shows the entire range of temperatures over the full transition duration, the right panel is zoomed over the temperature values between completion and nucleation.}
    \label{fig:physical_volume_fast}
\end{figure}

\begin{figure}[htb]
    \centering
    \includegraphics[width=0.99\linewidth]{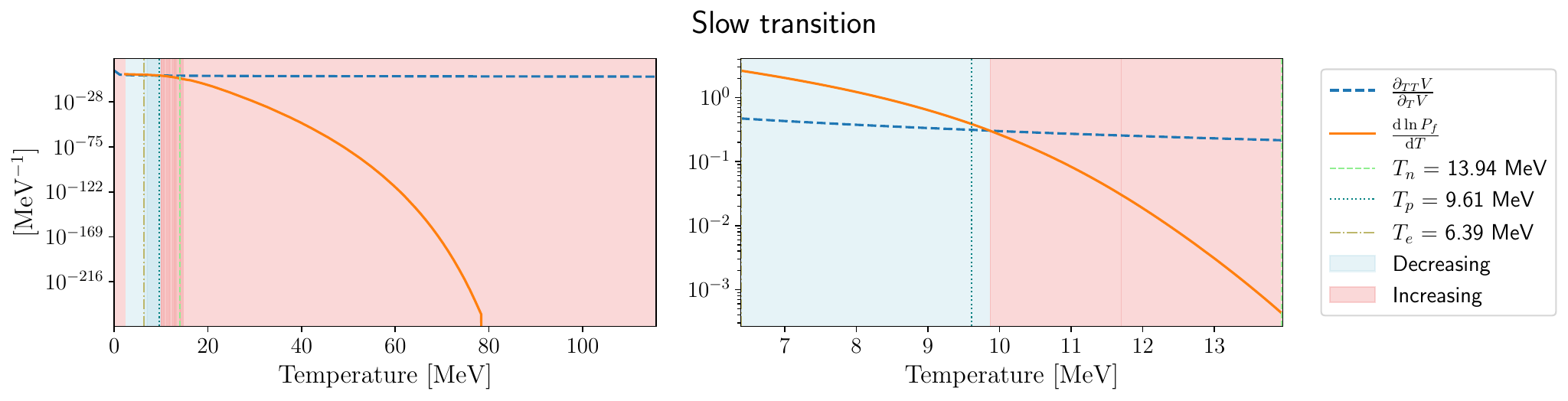}
    \caption{Same as in~\Cref{fig:physical_volume_fast} but for the ``Slow'' point in~\Cref{tab:input}.}
    \label{fig:physical_volume_slow}
\end{figure}

The evolution of the physical volume of the Universe in the false vacuum is reported in~\Cref{fig:physical_volume_fast} for the ``Fast'' transition, and in~\Cref{fig:physical_volume_slow} for the ``Slow'' one. In each figure the dashed curves represent the value of $\partial_{TT} V^\textrm{T} / \partial_T V^\textrm{T}$, while the continuos curves represent $\de \ln P_f / \de T$; the temperature values where the condition in~\Cref{eq:volume_decreasing} is satisfied (and thus where the physical volume in the false vacuum is decreasing) have a light blue background, while temperature values for which $\mathcal{V}_\textrm{phys}$ is increasing have a light coral background. For reference, the values of the milestone temperatures are reported as vertical dashed ($T_n$), dotted ($T_p$) and dot-dashed ($T_e$) lines. The left panels show the evolution of $\mathcal{V}_\textrm{phys}$ over the full range of temperatures $T \in [T_{min}, T_c]$, while the right panels are zoomed over $T \in [T_e, T_n]$.
It can be noticed that, for both benchmarks, $\mathcal{V}_\textrm{phys}$ is actually increasing at $T_n$, providing further evidence that the use of the nucleation temperature as the characteristic temperature for the computation of the SGWB spectrum should be avoided. In the ``Fast'' benchmark point $\de \ln P_f / \de T$ rapidly increases at low temperatures with respect to $\partial_{TT} V^\textrm{T} / \partial_T V^\textrm{T}$, signalling a fast conversion of the physical volume from false to true vacuum state.
In the ``Slow'' transition, the value of $\mathcal{V}_\textrm{phys}$ varies more gently, as evidenced by the smaller hierarchy of values between $\de \ln P_f / \de T$ and $\partial_{TT} V^\textrm{T} / \partial_T V^\textrm{T}$ when the physical volume in false vacuum is decreasing, as well as by the extended range of temperatures over which the transition takes place.
For both the ``Fast'' and ``Slow'' benchmark points in~\Cref{tab:input}, the condition~\Cref{eq:volume_decreasing} is satisfied at the percolation temperature.
Nevertheless, it could also be the case that $\mathcal{V}_{\textrm{phys}}$ decreases at $T_p$ but increases afterwards, which makes it fundamental to also check the completion condition from~\Cref{eq:condition_physical_volume}.

\subsection{Computation of thermal parameters determining the GW production}
The production of GW sourced by a FOPT mainly depends on three thermal parameters characterising the transition: the reheating temperature $T_\textrm{RH}$, the strength of the transition $\alpha$ and the mean bubble separation at percolation $R_*$~\cite{Athron:2023xlk}~\footnote{Note that the characteristic length-scale given by $R_*$ is sometimes traded for a characteristic duration of the FOPT $\beta^{-1}$, especially when computing gravitational waves generated by bubble collisions and relativistic shells~\cite{Jinno:2016vai,Lewicki:2022pdb}. The parameter $\beta$ can be related to $R_*$ through $\beta = (8\pi)^{1/3} v_w/R_*$, assuming the nucleation rate to follow $\Gamma\sim \Gamma_n e^{\beta t}$. Notice that slow transitions can significantly deviate from this assumption, as shown later in Section~\ref{sec:bubble_radius}.}. 
In addition, the efficiency parameters $\alpha_\infty$~\cite{Espinosa:2010hh} and $\alpha_{eq}$~\cite{Ellis:2019oqb} determine the fraction of  vacuum energy that is converted into GWs, as well as the distribution of this energy fraction into the different sources of GWs (bubble walls collisions, sound waves and turbulence in the plasma).
Finally, the transition strength and the spectral shape of the GWs produced by sound waves and turbulence depend as well on the speed of sound in the plasma, which in the presence of a non-negligible vacuum energy differs from the ideal relativistic gas assumed in the bag model. 

\elena implements general formulae for the computation of these quantities, going beyond common assumptions usually found in the literature. We exemplify how to compute the thermal parameters of a FOPT and review their expressions in the following subsections.

\subsubsection{Transition strength $\alpha$ and efficiency parameters $\alpha_{\infty}$, $\alpha_{eq}$}
Several definitions exist for the computation of the transition strength $\alpha$ (cf.~\cite{Athron:2023xlk} for a detailed analysis). \elena adopts the general definition~\cite{Giese:2020rtr,Giese:2020znk}
\begin{equation}
    \alpha = \frac{4}{3}\frac{\bar{\theta}_f(T_p)-\bar{\theta}_t(T_p)}{w_f(T_p)}\,,
    \label{eq:alpha_trace_anomaly}
\end{equation}
where $w = - T (\partial V^\textrm{T}/\partial T)$ is the enthalpy density of the full system (Standard Model plus new fields), $\bar{\theta} = (\rho-3p/c_{s,t}^2)/4$ is the pseudotrace, with $c_{s,t}^2(T)$ the speed of sound in the plasma in the broken phase, $\rho$ the energy density and $p$ the pressure. The pseudotrace is a generalisation of the vacuum energy. The latter should only be used when the assumptions of the MIT bag model are fulfilled~\cite{Athron:2023xlk}. In~\Cref{eq:alpha_trace_anomaly}, the subscript $f\,(t)$ indicates the value of these quantities in the false (true) minimum. This definition of $\alpha$ does not assume the validity of the bag model, preventing the appearance of large errors in the estimation of the kinetic energy fraction~\cite{Giese:2020rtr,Giese:2020znk}.

The value of $\alpha$ determines the total amount of energy that is released in the FOPT. The fraction of energy that is converted into each GW source depends on the efficiency parameter $\alpha_\infty$~\cite{Espinosa:2010hh}
\begin{equation}\label{eq:alpha_inf}
    \alpha_{\infty} = \frac{1}{18}\frac{\Delta m^2 T_p^2}{w_f(T_p)},\hspace{1cm} \,\mathrm{with}\, \hspace{0.5cm} \Delta m^2 = \sum_i c_i\ n_i(m_{i,t}^2-m_{i,f}^2),
\end{equation}
where the sum runs over all species gaining a mass, $n_i$ are the internal degrees of freedom and $c_i = 1\ (1/2)$ for bosons (fermions). The parameter $\alpha_{\infty}$ represents the minimal transition strength required to fully overcome the leading-order (LO) friction term of the plasma, which results from the fields becoming massive due to the symmetry breaking: if $\alpha > \alpha_\infty$, the excess energy goes into accelerating the bubble walls, resulting into bubble-wall velocities $v_w\rightarrow 1$. In the opposite scenario, the bubble walls reach a terminal velocity $v_w < 1$, whose exact determination requires a dedicated analysis~\cite{Espinosa:2010hh}. 

If the bubble walls reach relativistic velocities, next-to-leading order (NLO) friction terms, which are proportional to the wall Lorentz factor $\gamma_w$, become relevant. The effect of the NLO friction is instead quantified by the parameter $\alpha_\textrm{eq}$~\cite{Bodeker:2009qy, Bodeker:2017cim, Gouttenoire:2021kjv}, given by
\begin{equation}\label{eq:alpha_eq}
    \alpha_{\mathrm{eq}} = \frac{4g^2\Delta m_V T_p^3}{3w_f(T_p)}, \, \hspace{1cm} \mathrm{with}\, \hspace{0.5cm} g^2 \Delta m_V = \sum_X 3g_X^2 (m_{X,t}-m_{X,f}).
\end{equation}
The sum in~\Cref{eq:alpha_eq} only runs over the gauge bosons that become massive during the transition. The quantity $\alpha_\textrm{eq}$ is related to the terminal velocity that the bubble walls reach due to NLO friction terms~\cite{Bodeker:2017cim,Gouttenoire:2021kjv}
\begin{equation}\label{eq:gamma_eq}
    \gamma_{\mathrm{eq}}=\frac{\alpha-\alpha_{\infty}}{\alpha_{\mathrm{eq}}}\,.
\end{equation}
$\gamma_\textrm{eq}$ needs to be compared to the Lorentz factor that the wall would reach in the absence of the NLO pressure~\cite{Ellis:2019oqb}
\begin{equation}\label{eq:gamma_terminal}
    \gamma_* \equiv \frac{2}{3}\frac{R_*}{R_0}\,,
\end{equation}
where the parameter $R_0$ is defined in Ref.~\cite{Ellis:2019oqb}. Strictly speaking, the quantity appearing in the numerator of~\Cref{eq:gamma_terminal} should be the mean bubble radius. However, the mean bubble separation and mean bubble radius are comparable at percolation, and $R_*$ must in any case be computed to determine the SGWB spectrum, cf.~\Cref{sec:bubble_radius}. Thus, we can trade the two quantities for the sake of numerical efficiency, provided~\Cref{eq:gamma_terminal} is only evaluated at the percolation temperature.
Similarly to the LO case, if $\gamma_*\leq \gamma_{\mathrm{eq}}$, all the vacuum energy is then converted into accelerating the bubble walls, while if $\gamma_*> \gamma_{\mathrm{eq}}$, the bubble walls reach a terminal velocity and the exceeding energy shall be released into the surrounding plasma, generating GWs via sound waves and turbulence.

The quantities discussed in this subsection need to be derived only at the percolation temperature, which represents the moment when most of the GWs are generated. For completeness, we show in the~\Cref{code:alphas} how to evaluate them in \elena at any generic temperature $T \in [T_{min}, T_c]$ where the function \py{action_over_T(T)} from the~\Cref{code:action_over_T} has already been executed. The user can easily adapt the example presented in the~\Cref{code:alphas} to different models or computation needs.
\begin{lstlisting}[language=Python, caption=Computation of the transition strength and of the efficiency parameters., label=code:alphas]
from GWparams import alpha_th_bar # This is the definition of \alpha

def c_alpha_inf(T, units):
    v_true = true_vev[T]
    v_false = false_vev[T]
    Dm2_photon = 3 * g**2 * (v_true**2 - v_false**2)
    Dm2_scalar = 3 * lambda_ * (v_true**2 - v_false**2) 
    numerator = (Dm2_photon + Dm2_scalar) * T**2 / 24
    rho_tot = - T * 3 * (dp.dVdT(v_false, T, include_radiation=True, include_SM = True, units = units) ) / 4
    rho_DS = - T * 3 * (dp.dVdT(v_false, T, include_radiation=True, include_SM = False, units = units) ) / 4
    return numerator/ rho_tot, numerator / rho_DS

def c_alpha_eq(T, units):
    v_true = true_vev[T]
    v_false = false_vev[T]
    numerator = (g**2 * 3 * (g * (v_true - v_false)) * T**3)
    rho_tot = - T * 3 * (dp.dVdT(v_false, T, include_radiation=True, include_SM = True, units = units) ) / 4
    rho_DS = - T * 3 * (dp.dVdT(v_false, T, include_radiation=True, include_SM = False, units = units) ) / 4
    return numerator / rho_tot, numerator / rho_DS

action_over_T(T_perc)
alpha, alpha_DS = alpha_th_bar(T_perc, dp, V_min_value, false_vev, true_vev, units = units)
alpha_inf, alpha_inf_DS = c_alpha_inf(T_perc, units)
alpha_eq, alpha_eq_DS = c_alpha_eq(T_perc, units)

gamma_eq = (alpha - alpha_inf) / alpha_eq

if alpha < alpha_inf: # this implies v_w < 1
    is_physical = False

if gamma_eq < 1 / np.sqrt(1 - 0.99**2): # this would imply v_w < 0.99
    is_physical = False
\end{lstlisting}
\elena provides the function \py{GWparams.alpha_th_bar} to compute the transition strength at a given temperature using the~\Cref{eq:alpha_trace_anomaly}. The definitions of $\alpha_\infty$ and $\alpha_\textrm{eq}$ shall be provided by the user following the specific model under analysis; we provide in the~\Cref{code:alphas} the code to implement these computations for the model in~\Cref{eq:lag_UV}, defining the functions \py{c_alpha_inf} and \py{c_alpha_eq}, respectively. Notice that, while the masses appearing in~\Cref{eq:alpha_inf,eq:alpha_eq} generally have a vacuum component (function of the scalar vevs) and a thermal component (function of the temperature), the latter one cancels in the differences as long as the temperature on both sides of the bubble wall is the same. The functions \py{GWparams.alpha_th_bar}, \py{c_alpha_inf} and \py{c_alpha_eq} return each a tuple of two elements: the first one is the result of~\Cref{eq:alpha_trace_anomaly}, while the second one is the transition strength normalised to the enthalpy density of the non-Standard Model fields only, i.e. $w_\textrm{BSM} = - T (\partial V^\textrm{BSM}/\partial T)$. These further efficiency factors are relevant to the transition dynamics in scenarios where the new fields are effectively decoupled from the SM, forming a secluded dark sector~\cite{Ertas:2021xeh}.

The functions \py{c_alpha_inf} and \py{c_alpha_eq} in the~\Cref{code:alphas} simply take as argument the temperature \py{T}, expressed in the appropriate \py{units}. The function \py{GWparams.alpha_th_bar} takes as arguments the temperature (\py{T_perc} in this example), the model class (\py{dp}) and three dictionaries \py{V_min_value}, \py{false_vev}, and \py{true_vev}, each containing for the key temperature the values of the free energy, and the location of the false and true vacua, respectively. The line \py{action_over_T(T_perc)} computes these quantities at the temperature \py{T_perc} (cf.~\Cref{code:action_over_T}). The following lines compute the transition strength normalised to the total enthalpy density (\py{alpha}) as well as to the BSM contribution only (\py{alpha_DS}), the LO  (\py{alpha_inf}, \py{alpha_inf_DS}) and NLO efficiency factors (\py{alpha_eq}, \py{alpha_eq_DS}) with analogous normalisation conventions. \py{gamma_eq} is the resulting terminal bubble wall Lorentz factor,~\Cref{eq:gamma_eq}.
Some discussion is in order concerning the last block of code in~\Cref{code:alphas}: here the boolean flag \py{is_physical} is set to \py{False} if the condition $\alpha > \alpha_\infty$ is not met, or if $\gamma_{\mathrm{eq}}\lesssim7$ (corresponding to a wall velocity $v_w < 0.99$). Even though these solutions are perfectly viable from a FOPT point of view, the computation of  $P_f (T)$ previously performed is not reliable in these cases, given that the argument \py{v_w} was set to \py{1} in \py{temperatures.compute_logP_f} (cf.~\Cref{code:log_Pf}). As previously mentioned, in the case $\alpha < \alpha_\infty$, a dedicated analysis is necessary to estimate the terminal bubble wall velocity $v_w$. \elena can compute the transition history for generic values of the bubble wall velocity by setting the appropriate argument \py{v_w}, but the physical value of $v_w < 1$ shall be provided by the user. Since in the example code provided here we assumed $v_w = 1$, the flag \py{is_physical} is then set to \py{False} if the transition is not strong enough to reach $v_w \rightarrow 1$, signalling a breaking of the underlying assumptions and the need to explicitly compute the value of $v_w$. This should however not be interpreted as an exclusion of the considered benchmark point.

\begin{figure}[htb]
    \centering
    \includegraphics[width=0.99\linewidth]{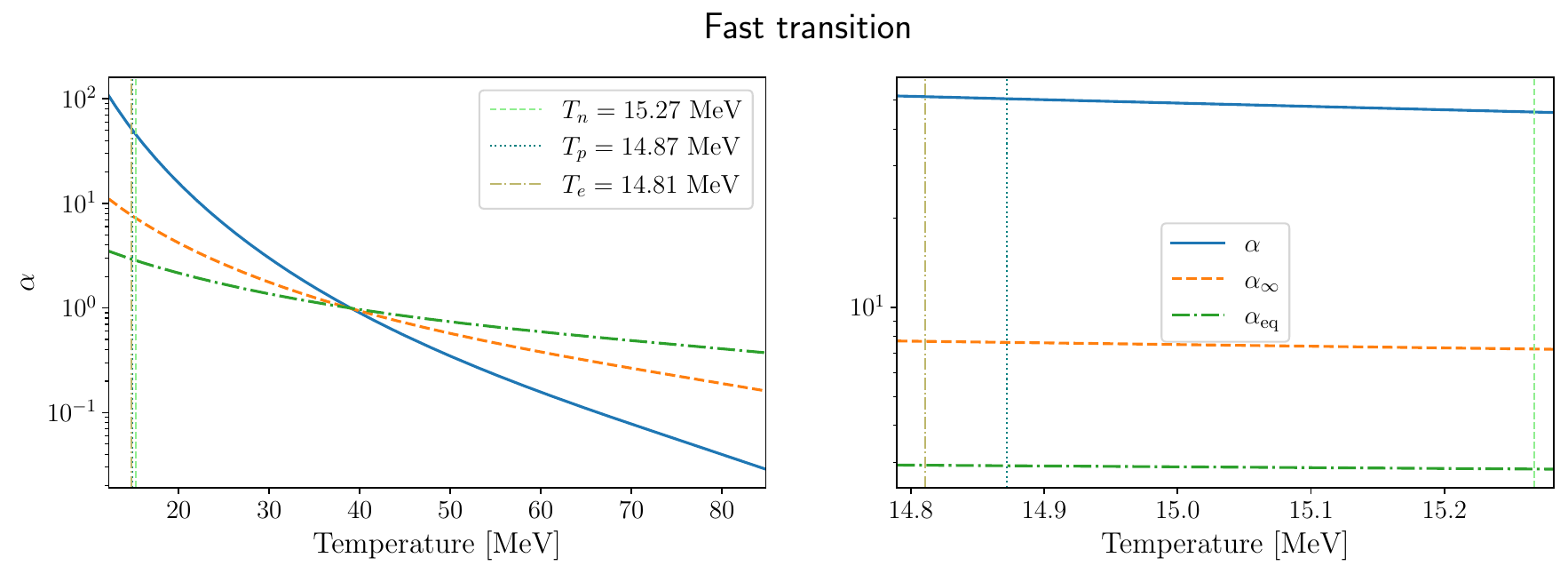}
    \caption{Evolution of the available transition strength $\alpha$ (continuos curve), LO friction term $\alpha_\infty$ (dashed curve) and NLO friction term (dot-dashed curve) as function of temperature for the ``Fast'' benchmark point in~\Cref{tab:input}. The nucleation ($T_n$), percolation ($T_p$) and completion ($T_e$) temperatures are signalled as vertical dashed, dotted and dot-dashed lines, respectively. The left panel shows the full temperature range over the transition evolution, while the right panel zooms on the interval between $T_e$ and $T_n$.}
    \label{fig:alpha_fast}
\end{figure}

\begin{figure}[htb]
    \centering
    \includegraphics[width=0.99\linewidth]{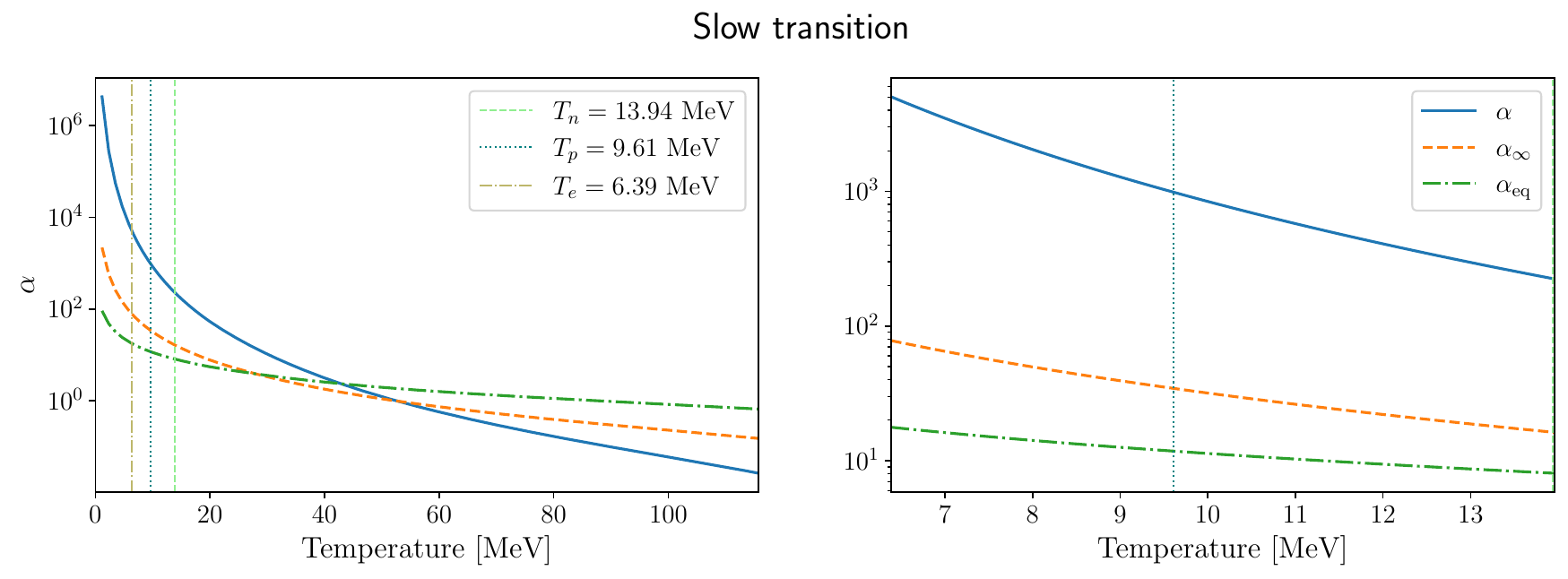}
    \caption{Same as in~\Cref{fig:alpha_fast} but for the ``Slow'' benchmark in~\Cref{tab:input}.}
    \label{fig:alpha_slow}
\end{figure}

We report the evolution of $\alpha$ (continuos curves), $\alpha_\infty$ (dashed curves) and $\alpha_\textrm{eq}$ (dot-dashed curves) as functions of temperature in~\Cref{fig:alpha_fast} (\Cref{fig:alpha_slow}) for the ``Fast'' (``Slow'') benchmark in~\Cref{tab:input}. Left panels show the full temperature range $T \in [T_{min}, T_c]$, while right panels are zoomed over $T \in [T_e, T_n]$. For reference, the nucleation ($T_n$), percolation ($T_p$) and completion ($T_e$) temperatures are signalled as vertical dashed, dotted and dot-dashed lines, respectively.
We notice that in both benchmark points the condition $\alpha > \alpha_\infty$ is satisfied at $T_n$, which is when bubble nucleation starts to be relevant, thus justifying the assumption on the bubble walls velocity $v_w = 1$ in the computation of $P_f$.

\subsubsection{Mean bubble separation and transition duration}\label{sec:bubble_radius}
\begin{figure}[htb]
    \centering
    \includegraphics[width=0.99\linewidth]{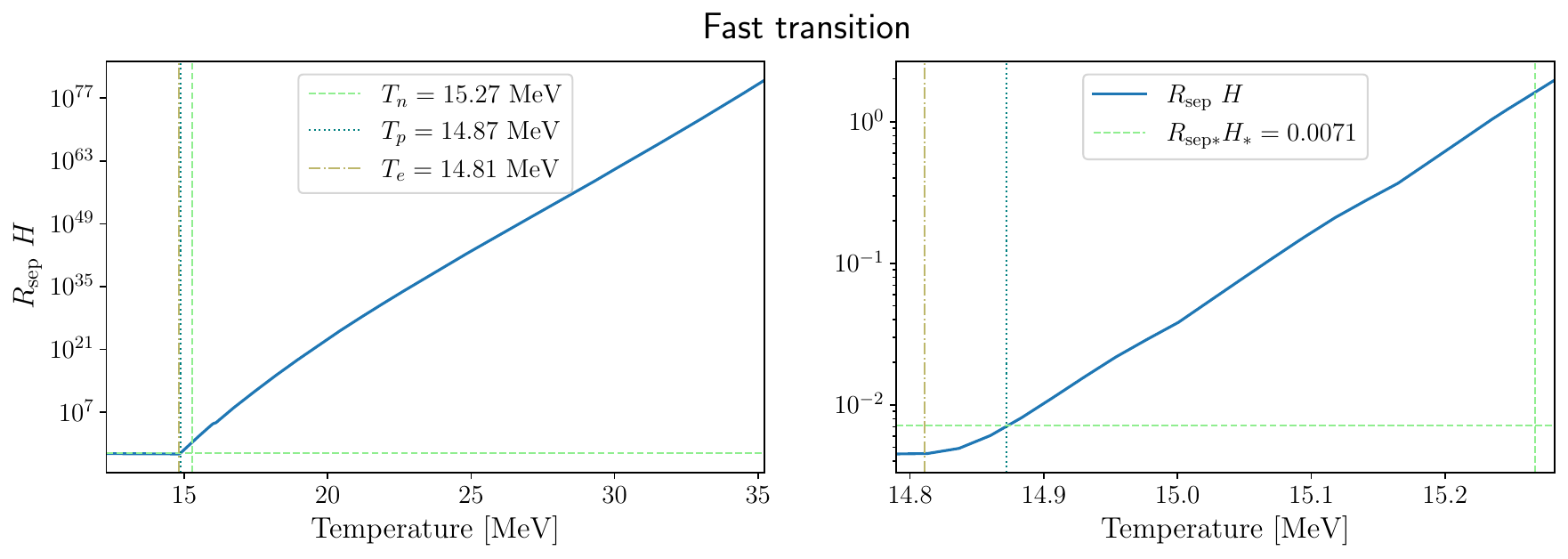}
    \caption{Evolution of the mean bubble separation in units of the Hubble radius, $R_\textrm{sep}(T) H(T)$, as a function of temperature for the ``Fast'' benchmark point in~\Cref{tab:input}. Vertical lines represent the nucleation ($T_n$), percolation ($T_p$) and completion ($T_e$) temperatures as dashed, dotted and dot-dashed lines, respectively. The horizontal dashed line represents the value of the mean bubble separation at percolation $R_\textrm{sep} (T_p) H(T_p) = R_\textrm{sep}{}_* H_*$. The left panel shows the evolution over the full range of temperatures, the right panel zooms on the interval between $T_n$ and $T_e$.}
    \label{fig:RH_fast}
\end{figure}

\begin{figure}[htb]
    \centering
    \includegraphics[width=0.99\linewidth]{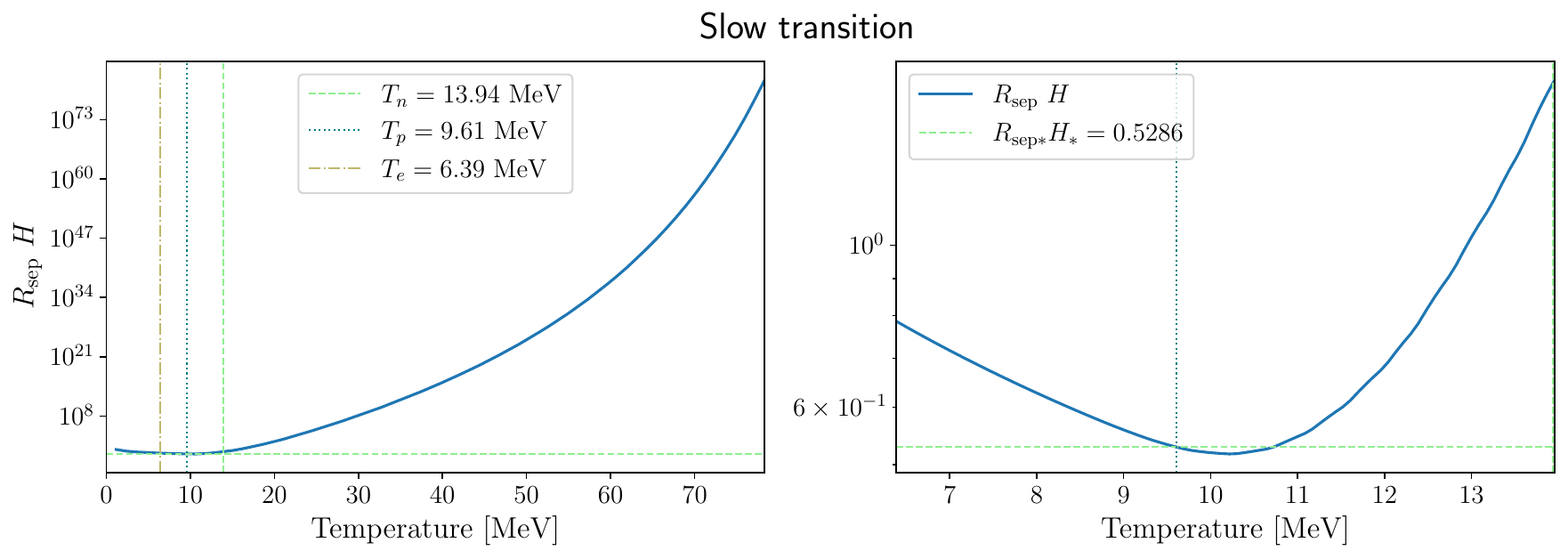}
    \caption{Same as in~\Cref{fig:RH_fast}, but for the ``Slow'' benchmark point in~\Cref{tab:input}.}
    \label{fig:RH_slow}
\end{figure}

The amplitude and peak frequency of the SGWB produced by a FOPT strongly depend on the characteristic length-scale of the transition, which can be specified using different parameters, requiring to choose one among them~\cite{Athron:2023xlk}. The parameter used as input in many numerical hydrodynamic simulations that provide fits to the SGWB spectrum (cf. e.g.~\cite{Hindmarsh:2015qta}) is the mean bubble separation
\begin{equation}
    R_\textrm{sep} \left(t\right) = n \left(t\right)^{-\frac{1}{3}},
\end{equation}
where $n$ is the bubble number density~\cite{Guth:1979bh} given by
\begin{equation}\label{eq:bubble_number_density}
    n \left( t \right) = \int_{t_c}^t \de t' \Gamma \left( t' \right)  P_f \left( t' \right) \frac{a^3\left( t' \right)}{a^3\left( t \right)}.
\end{equation}
Given that the use of $R_\textrm{sep}$ is recommended over other possible length-scale choices~\cite{Athron:2023xlk}, \elena implements the computation of the mean bubble separation with the function \py{temperatures.R_sepH}, which returns a tuple of two floats, $R_\textrm{sep}(T) H(T)$ and $R_\textrm{sep}$. The first quantity normalises the mean bubble separation to the Hubble radius, and is the one inputted in SGWB fits; the second quantity is the dimensionful physical distance. The use of \py{temperatures.R_sepH} is exemplified in the~\Cref{code:RH}.
\begin{lstlisting}[language=Python, caption=Computation of the mean bubble separation., label=code:RH]
from temperatures import R_sepH

RH, R = R_sepH(Temps, Gamma, logP_f, H, ratio_V)
RH_interp = interpolation_narrow(Temps, RH, T_perc)
\end{lstlisting}
The function \py{temperatures.R_sepH} requires as arguments the outputs of \py{temperatures.compute_logP_f}, see the~\Cref{code:log_Pf} and subsequent discussion. The arrays \py{RH} and \py{R} contain the value of $R_\textrm{sep}(T) H(T)$ and $R_\textrm{sep}(T)$ for each temperature value in the array \py{Temps}. In the last line of~\Cref{code:RH}, the function \py{utils.interpolation_narrow} is used to find the value of the mean bubble separation at the percolation temperature, \py{RH_interp}.
We report as a continuous curve in~\Cref{fig:RH_fast} (\Cref{fig:RH_slow}) the evolution of $R_\textrm{sep}(T) H(T)$ for the ``Fast'' (``Slow'') benchmark point in~\Cref{tab:input}. The temperatures of nucleation ($T_n$), percolation ($T_p$) and completion ($T_e$) are reported as vertical dashed, dotted and dot-dashed lines, respectively. The horizontal dashed line represents the mean bubble separation at percolation $R_\textrm{sep} (T_p) H(T_p) = R_\textrm{sep}{}_* H_*$, which is the parameter entering into the SGWB spectrum. Left panels show the full range of transition temperatures, while right panels provide a zoomed view over the temperature range between nucleation and completion.
For the ``Fast'' transition in~\Cref{fig:RH_fast} the nucleation of bubbles is sizeable until the completion temperature $T_e$, thus resulting in a relatively small value of the mean bubble separation at percolation in units of the Hubble radius, $(R_\textrm{sep}{}_* H_*)^\textrm{fast} = 7.1 \times 10^{-3}$.
For the ``Slow'' transition in~\Cref{fig:RH_slow}, on the other hand, the nucleation of new bubbles becomes ineffective before the percolation temperature $T_p$, and the conversion of volume from false to true vacuum mainly proceeds via the expansion of already nucleated bubbles. This results in a slow increase in $R_\textrm{sep}(T) H(T)$ due to the expansion of the Universe, as expected from~\Cref{eq:bubble_number_density}, even though the physical volume in the false vacuum is decreasing due to the bubbles walls expansion, cf.~\Cref{fig:physical_volume_slow}. The resulting mean bubble separation at percolation in units of Hubble radius is thus sizeable, $(R_\textrm{sep}{}_* H_*)^\textrm{slow} = 0.53$.

An alternative to $R_\textrm{sep}$ widely used in the literature is the inverse time-scale of the transition, $\beta$~\cite{Athron:2023xlk}.
This quantity is related to the shape of the tunnelling action $S_{E,d}$ when it can be Taylor expanded around the nucleation time and truncated to the linear term as
\begin{equation}
    \beta_* = - \left. \frac{\de S}{\de t}\right|_{t_*}.
\end{equation}
This computation thus assumes an exponential nucleation rate, $\Gamma(t) = \Gamma(t_*) \exp(\beta_* (t-t_*))$. However, an exponential nucleation rate is characteristic of fast transitions only, while using this parametrisation for slow transitions can give wrong results. We show this for our benchmark points in~\Cref{tab:input} in the following, by employing the more general parametrisation for the nucleation rate that includes the quadratic term in the Taylor expansion 
\begin{equation}\label{eq:taylor2}
    \Gamma(t) = \Gamma_n \exp\left[ \beta (t - t_n) - \frac{1}{2} \gamma^2 (t - t_n)^2 \right],
\end{equation}
where $t_n$ is the nucleation time and $\Gamma_n = \Gamma(t_n)$ is the false vacuum decay rate at nucleation. 
Time can be related to temperature by
\begin{equation}\label{eq:T_to_t}
    t - t_0 = \int_{t_0}^t \de t = \int_{T\left(t\right)}^{T\left(t_0\right)} \frac{\de T'}{3 H\left(T'\right)} \frac{\partial_{TT} V^\textrm{T} \left(T'\right)}{\partial_T V^\textrm{T} \left(T'\right)}.
\end{equation}
\elena provides the function \py{GWparams.beta} to readily extract the values of the $\beta$ and $\gamma$ parameters in~\Cref{eq:taylor2}, by simply using as arguments the output of \py{temperatures.compute_logP_f}(cf.~\Cref{code:milestones}), a starting (\py{T_nuc}) and final (\py{T_perc}) temperatures, as well as an optional argument \py{verbose} (default \py{False}) that controls its output.
To avoid the appearance of dimensionful quantities that can be very small and result in numerical instabilities, \py{GWparams.beta} normalises time in units of the Hubble expansion rate at nucleation $H_n$, thus employing $H_n (t-t_n)$ as the independent variable, and extracting $\beta/H_n$ and $\gamma / H_n$ as coefficients. The usage of \py{GWparams.beta} is demonstrated in the~\Cref{code:nucleation}.
\begin{lstlisting}[language=Python, caption=Computation of the nucleation rate coefficients., label=code:nucleation]
from GWparams import beta

beta_Hn, gamma_Hn, times, Gamma_t, Temps_t, H_t = beta(Temps, ratio_V, Gamma, H, T_nuc, T_perc, verbose = True)
\end{lstlisting}
The function \py{GWparams.beta} performs a fit of the numerically computed decay rate $\Gamma(t)$ to the expression in~\Cref{eq:taylor2}, setting $t_n$ as the time of the starting temperature, and considering a time interval extended until the final temperature. If \py{verbose = False} the function only returns the values of $\beta/H_n$ (\py{beta_Hn}) and $\gamma / H_n$ (\py{gamma_Hn}). If \py{verbose = True} the function also returns an ensemble of arrays, containing in each element the corresponding values of $H_n \times (t-t_n)$ (\py{times}), $\Gamma(t - t_n)$ (\py{Gamma_t}), $T(t - t_n)$ (\py{Temps_t}) and $H(t - t_n)$ (\py{H_t}), where the $Q(t-t_n)$ notation indicates that the $Q$ quantity is evaluated at different ($t - t_n$) time values~\footnote{We write instead $H_n \times (t-t_n)$ to stress that this is a multiplication between the constant quantity $H_n$  and the time variable $t - t_n$.}.

\begin{figure}[htb]
    \centering
    \includegraphics[width=0.99\linewidth]{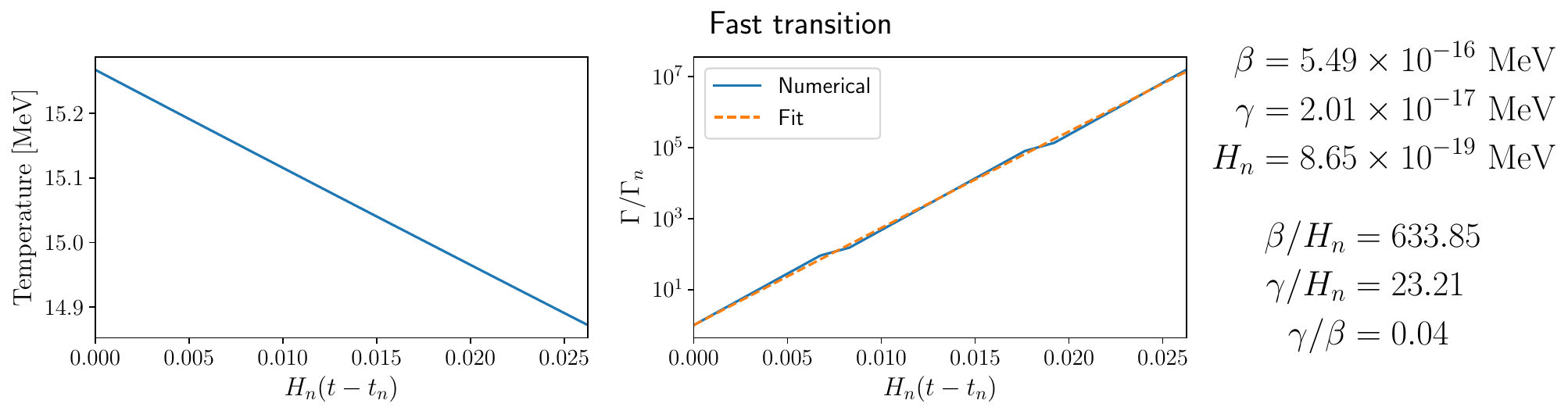}
    \caption{
    Nucleation evolution for the ``Fast'' point in~\Cref{tab:input}. The left panel shows the evolution of temperature as a function of time between nucleation and percolation, with the time expressed in units of the Hubble time at nucleation $1/H_n$. The right panel shows the nucleation rate computed numerically in \elena (continuos line) and the result of~\Cref{eq:taylor2} with the numerical parameters reported on the right side of the figure (dashed line).
    }
    \label{fig:nucleation_fast}
\end{figure}

\begin{figure}[htb]
    \centering
    \includegraphics[width=0.99\linewidth]{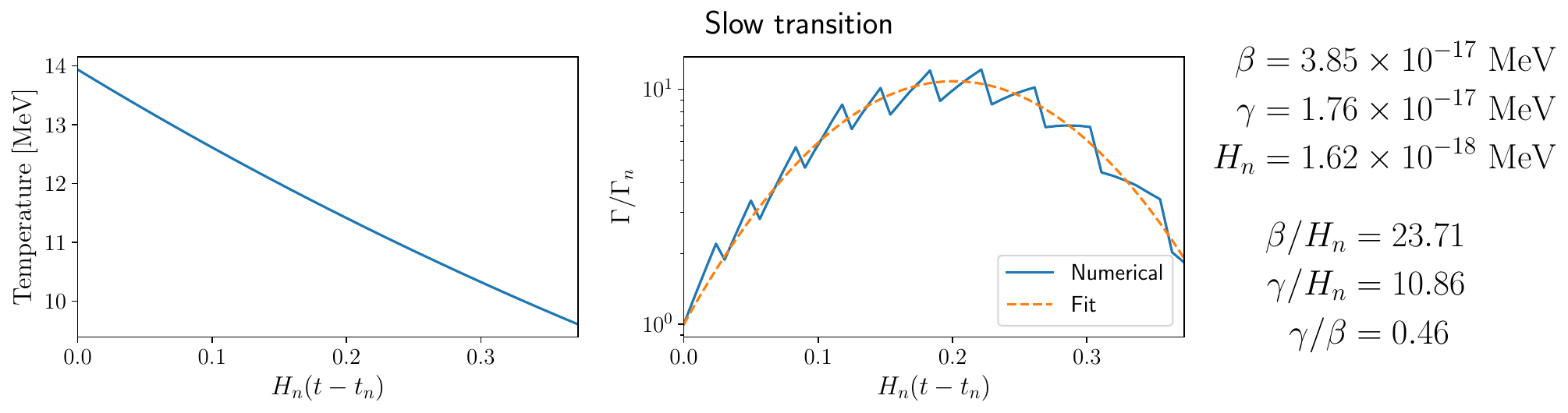}
    \caption{Same as in~\Cref{fig:nucleation_fast} but for the ``Slow'' point in~\Cref{tab:input}.}
    \label{fig:nucleation_slow}
\end{figure}

The results of the numerical fit to~\Cref{eq:taylor2} for the ``Fast'' point in~\Cref{tab:input} are reported in~\Cref{fig:nucleation_fast} for the evolution between the nucleation and percolation times. The left panel in the figure shows the evolution of the temperature with time, from which it can be noted that percolation is reached after $2.6 \times 10^{-2}$ unit times $1/H_n$. On the right panel we report the numerical computation of the nucleation rate (continuous line) and the expression computed by inserting the fit results (reported on the right-side of the figure) into~\Cref{eq:taylor2}. Clearly, the exponential nucleation is a very good approximation for this point, with the ratio $\gamma / \beta = 0.04$ indicating that the second-order term in the Taylor expansion can be safely neglected. In this case, using the $\beta$ parameter to characterise the transition is justified.
Analogously, we report in~\Cref{fig:nucleation_slow} the results of the numerical fit to the ``Slow'' benchmark point in~\Cref{tab:input}. As can be noted in the left panel and as expected, the duration of this transition is much longer than the ``Fast'' one, taking $0.37 / H_n$ from nucleation to percolation. More importantly, the results of the numerical fit reported on the right-side of the figure show that the second order term in the Taylor expansion is relevant, with a ratio $\gamma / \beta = 0.46$. Thus, this benchmark point cannot be approximated with an exponential nucleation, showing explicitly that the employment of $\beta$ to characterise the SGWB spectrum would produce unreliable results. 
As exemplified by the discussion in this Section, it is strongly advised to employ $R_\mathrm{sep}$ to characterise the typical length-scale of the transition~\cite{Athron:2023xlk}. This choice is followed in \elena, which provides the function \py{temperatures.R_sepH} for the computation of $R_\mathrm{sep}$; the user can nevertheless easily compute the $\beta$ and $\gamma$ parameters using \py{GWparams.beta}, if these quantities are of interest.

\subsection{Computation of the gravitational waves spectrum}
Once the thermal parameters of the FOPT are known, the SGWB can be finally computed. Among the different possibilities present in the literature, \elena employs the relations provided in Ref.~\cite{Ellis:2019oqb}, which use the mean bubble separation at percolation $R_\textrm{sep}{}_* H_*$ as input parameter (instead of the more commonly used inverse time-scale $\beta$) and that can thus also apply to scenarios where the nucleation rate is not exponential (cf. discussion in~\Cref{sec:bubble_radius}).

The GW spectra from Ref.~\cite{Ellis:2019oqb} are implemented in the class \py{GWparams.GW_SuperCooled}. The user can create an instance of the class by inputting the thermal parameters and then access the predicted GW spectra at any given frequency using the class methods, as exemplified in~\Cref{code:GW_spectra}.
\begin{lstlisting}[language=Python, caption=Computation of the stochastic gravitational waves background spectra., label=code:GW_spectra]
from GWparams import GW_SuperCooled

inst = GW_SuperCooled(T_perc, alpha, alpha_inf, alpha_eq, R_star, gamma_star, H_star, c_s = np.sqrt(c_s2), v_w = 1, units = units, dark_dof = 4)

log_freq_min, log_freq_max = -10, 10
x = np.linspace(log_freq_min, log_freq_max, 100)
x = 10**x

plt.plot(x, inst.Omegah2(x), label = "Total")
plt.plot(x, inst.Omegah2coll(x), label = "Collision")
plt.plot(x, inst.Omegah2sw(x), label = "Sound waves")
plt.plot(x, inst.Omegah2turb(x), label = "Turbulence")
plt.legend()
plt.plot()
\end{lstlisting}
In~\Cref{code:GW_spectra} a constructor call is used to create an instance \py{inst} of the class \py{GWparams.GW_SuperCooled}.
The constructor requires as mandatory arguments the percolation temperature $T_p$ (\py{T_perc}), the efficiency factors  $\alpha$, $\alpha_\infty$ and $\alpha_\textrm{eq}$  (\py{alpha}, \py{alpha_inf} and \py{alpha_eq}, respectively), the mean bubble separation at percolation $R_\textrm{sep}{}_*$ (\py{R_star}), the factor $\gamma_*$ (\py{gamma_star}), and the Hubble parameter at percolation $H(T_p)$ (\py{H_star}). The constructor also accepts optional arguments for the sound speed in the plasma at percolation $c_{s,t}(T_p)$ (\py{c_s}, default \py{1/np.sqrt(3)}), the bubbles walls speed (\py{v_w}, default \py{1}), the units of the dimensionful quantities (\py{units}, default \py{'GeV'}) and the relativistic degrees of freedom additional to the SM ones (\py{dark_dof}, default \py{4}).

Once the instance \py{inst} of \py{GWparams.GW_SuperCooled} has been constructed, the value of the SGWB density today $h^2 \Omega_\textrm{GW}(f)$ for each frequency $f$ (in Hertz) can be accessed by simply calling the method \py{GWparams.GW_SuperCooled.Omegah2}, e.g. \py{inst.Omegah2(1e-9)} to obtain the SGWB at $f=10^{-9}$ Hz. In addition, the contributions from each individual GW source can be accessed as well with the methods \py{GWparams.GW_SuperCooled.Omegah2coll}, \py{GWparams.GW_SuperCooled.Omegah2sw} and \py{GWparams.GW_SuperCooled.Omegah2turb}, for the SGWB sourced by bubbles walls collisions, sound waves and turbulence, respectively. In the~\Cref{code:GW_spectra} the class \py{GWparams.GW_SuperCooled} is used to plot the SGWB spectrum, but the user can take advantage of the modular structure of \elena to implement this computation in their own software or statistical inference tools.

\begin{figure}[htb]
    \centering
    \includegraphics[width=0.49\linewidth]{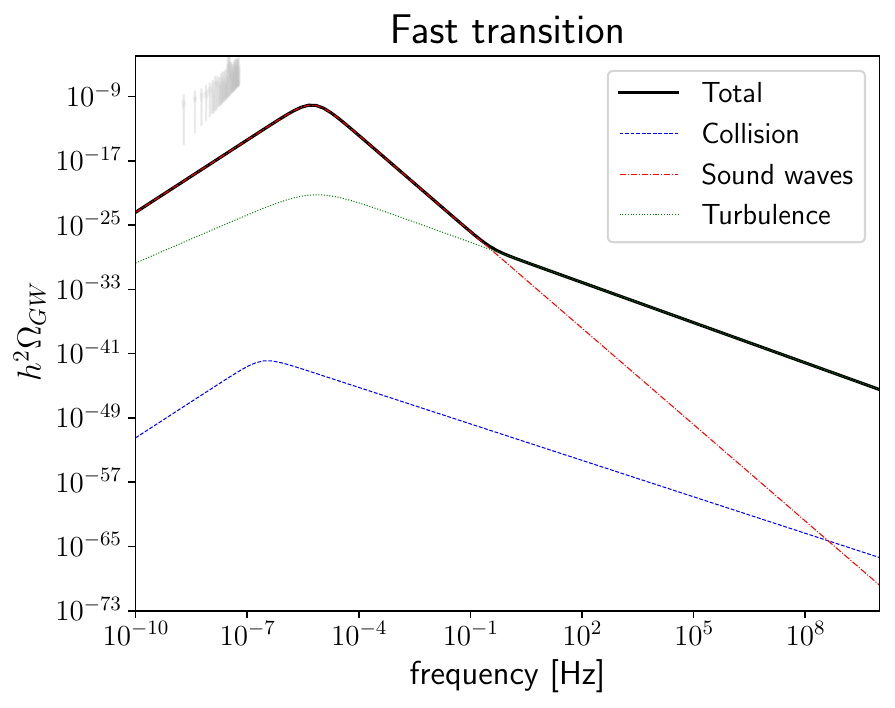}
    \includegraphics[width=0.49\linewidth]{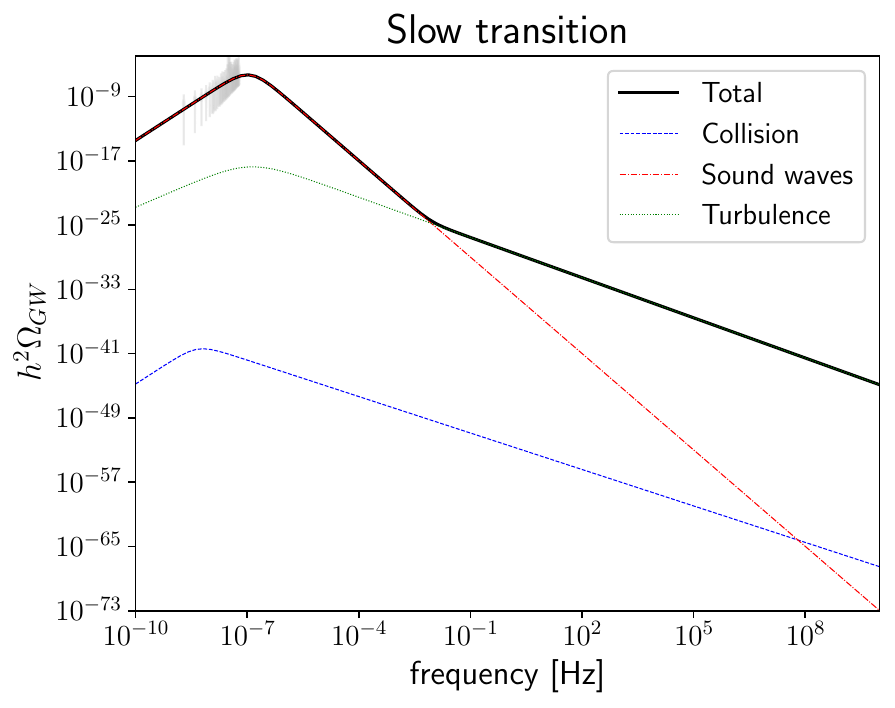}
    \caption{SGWB spectra predicted for the ``Fast'' (left panel) and ``Slow'' (right panel) benchmark points in~\Cref{tab:input}. Dashed, dot-dashed and dotted lines represent the individual contributions from bubbles walls collisions, sound waves and turbulence, while the continuos black lines are their sum. The gray bands represent the periodogram from the NANOGrav 15-years Dataset~\cite{NANOGrav:2023gor,NANOGrav:2023hde,the_nanograv_collaboration_2023_10344086}.}
    \label{fig:GW_spectra}
\end{figure}

We report in~\Cref{fig:GW_spectra} the SGWB spectra for the two benchmark points in~\Cref{tab:input}, together with the periodogram resulting from the NANOGrav 15-years Dataset~\cite{NANOGrav:2023gor,NANOGrav:2023hde,the_nanograv_collaboration_2023_10344086}.
As expected, the ``Slow'' transition results in a much stronger SGWB spectrum, peaked at lower frequencies with respect to the ``Fast'' one. Only the former can provide a viable solution to the NANOGrav data, with the latter producing a SGWB spectrum which is orders of magnitude smaller in amplitude at the relevant frequencies. 

We stress that this is only an example application of the code, and that \elena can compute the FOPT parameters for general scenarios, not necessarily related to Pulsar Timing Array observatories.

\section{Use case: MCMC analysis of a model using PTArcade}\label{sec:MCMC}
In this Section, we show an explicit use case that takes advantage of the speed and modularity of \elena to perform a statistical analysis on the parameters of a new physics model from observational data.
We interface \elena with \textsf{PTArcade}~\cite{Mitridate:2023oar} to perform a Markov Chain Monte Carlo (MCMC) analysis of the dark sector model from~\Cref{eq:lag_UV}, introduced in Ref.~\cite{Costa:2025csj}, comparing its predictions for the SGWB to the NANOGrav 15-year data. Notice that the model in~\Cref{eq:lag_UV} needs to be extended to also include portals to the SM sector, to ensure both thermalisation of the dark sector and the decay of its states into SM particles after the phase transition, guaranteeing compatibility with cosmological constraints~\cite{Bringmann:2023opz}. In the present analysis, given the scope of the work on the software package \elena, we focus for simplicity on the minimal dark sector in~\Cref{eq:lag_UV}, assuming that the additional interactions with the SM do not significantly affect the FOPT dynamics. We defer the analysis of extensions of~\Cref{eq:lag_UV} to future work, while stressing that the complications from considering a complete model are confined to the particle and cosmology phenomenology sides (e.g. computing the full thermal potential, the dark sector decay widths, or applying the experimental and observational constraints). The pipeline for computing the FOPT thermal parameters with \elena is unchanged; the user just needs to provide the correct model class for the specific thermal potential. Since a detailed discussion on possible extensions of~\Cref{eq:lag_UV} goes beyond the scope of the present work, we only apply a lower bound on the final reheating temperature, $T_\textrm{reh} \ge 4$ MeV~\cite{Hannestad:2004px}, where we estimate~\cite{Ellis:2018mja}
\begin{equation}
    T_\textrm{reh} = \left(1 + \alpha\right)^\frac{1}{4}\ T_p.
\end{equation}

\begin{figure}[htb]
    \centering
    \includegraphics[width=0.80\linewidth]{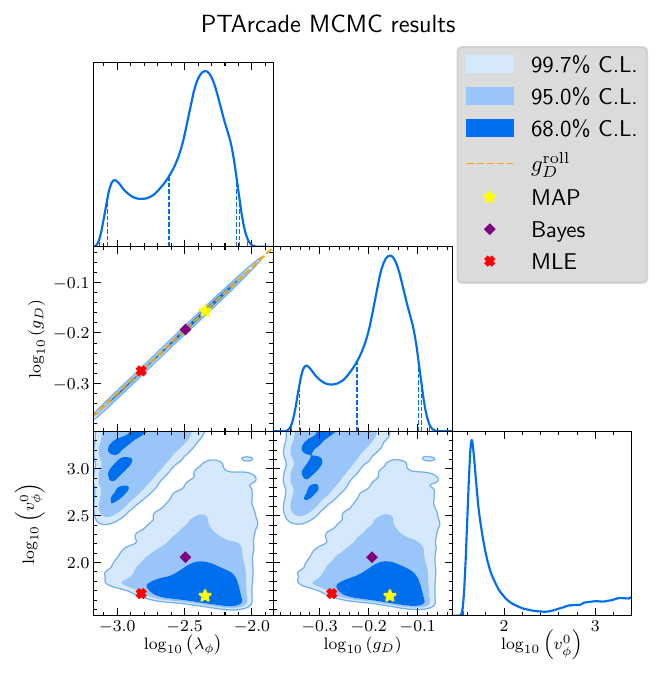}
    \caption{Posteriors for the model parameters in~\Cref{eq:lag_UV} derived from a MCMC study on the NANOGrav 15-year data, performed by interfacing \elena with \textsf{PTArcade}. We report the regions corresponding to 68\% (bright blue), 95\% (sky blue) and 99.7\% (pale blue) confidence levels, together with the maximum a posterior (MAP), maximum likelihood (MLE) and Bayes estimators of the three parameters as a yellow star, red cross and purple rhombus, respectively. For reference, we also plot $g_D^\textrm{roll}$ in~\Cref{eq:g_roll} as a dashed orange line.}
    \label{fig:MCMC}
\end{figure}

We use \textsf{PTArcade} to collect more than 9 millions samples\footnote{The final sample results from the union of 1340 individual chains.} (9,360,965) on the model in~\Cref{eq:lag_UV}: this is done by creating a script that contains the code developed in~\Cref{sec:software_chain} and using the final instance of the class \py{GW_SuperCooled} as input for the \py{spectrum} function required by \textsf{PTArcade}. For the parameter points where the variable \py{is_physical} results to \py{False} we assume no production of SGWB (i.e. $h^2 \Omega_\textrm{GW} = 0$ for all frequencies).
In Ref.~\cite{Costa:2025csj} it was shown that, in order to produce a strong FOPT, the full scalar potential generated from the Lagrangian in~\Cref{eq:lag_UV} must have a conformal-like shape, which is realised if the $g_D$ coupling lies close to the line defined by
\begin{equation}
g_{D}^{\mathrm{roll}} = \left\lbrace \frac{16\pi^2\lambda_{\phi}}{3}\left[1-\frac{\lambda_{\phi}}{8\pi^2}\left(5+2\log{2}\right)\right]\right\rbrace^{1/4}\,.
\label{eq:g_roll}
\end{equation}
To speed-up the convergence of the algorithm, instead of sampling with a uniform prior over the plane $(\log_{10}\lambda_\phi, \log_{10} g_D)$, we assume a uniform probability distribution over the line $g_D^\textrm{roll}$, and a gaussian distribution (centred at $g_D^\textrm{roll}$) over the perpendicular direction. This allows to reduce the burn-in phase, by aligning the sampling directions with the high-likelihood region. In terms of the Lagrangian parameters, the sampling covered the range of values: $\log_{10} \lambda_\phi \in (-3.17, -0.52)$, $\log_{10} g_D \in (-0.67, 0.62)$, $\log_{10} v_\phi^0 / \textrm{MeV} \in (1.0, 3.4)$ (with a uniform prior distribution for the latter).
The results of the analysis, rotated in the physical parameters, are reported in~\Cref{fig:MCMC}, where it is evident that the data strongly prefer a conformal-like potential (i.e. having $\lambda_\phi$ and $g_D$ lying around $g_D^\textrm{roll}$) with a new-physics scale below 100 MeV. The maximum a posterior (MAP), Bayes and maximum likelihood (MLE) estimators for the three parameters are identified by a yellow star, a purple rhombus and a red cross, respectively.
The MAP, MLE and Bayes estimators for the Lagrangian parameters are reported in~\Cref{tab:bayes}.
\begin{table}[]
    \centering
    \begin{tabular}{|c|c|c|c|}
    \hline
        Parameter & MAP & Bayes & MLE \\
        \hline
         $\log_{10} \lambda_\phi$ & -2.35 & $-2.49 \pm 0.28$ & -2.82 \\
         $\log_{10} g_D$ & -0.16 & $-0.19 \pm 0.07$ & -0.28 \\
         $\log_{10} v_\phi^0$ / MeV & 1.64 & $2.06 \pm 0.57$ & 1.67 \\
    \hline
    \end{tabular}
    \caption{Maximum a posterior (MAP), Bayes and maximum likelihood (MLE) estimators for the Lagrangian parameters in~\Cref{eq:lag_UV} from the NANOGrav 15-year dataset, obtained from the MCMC sampling performed with \textsf{PTArcade}, as described in the main text.
    }
    \label{tab:bayes}
\end{table}

\begin{figure}[htb]
    \centering
    \includegraphics[width=0.99\linewidth]{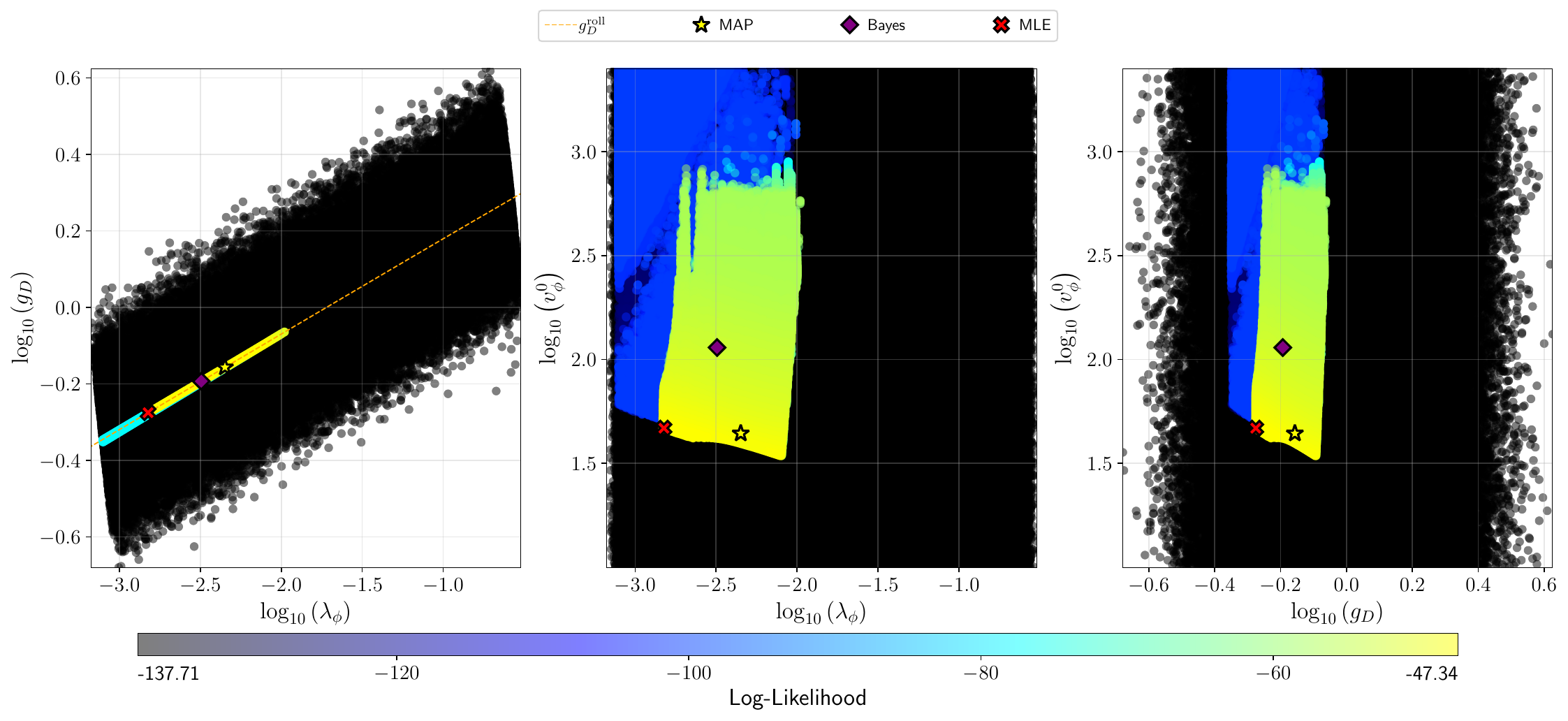}
    \caption{Scatter plots showing the sampled parameter space in two-dimensional slices. The colour bar corresponds to the value of the log-likelihood for each point. Yellow regions provide a larger probability of observing the NANOGrav 15-year signal given the chosen parameters. Black points correspond to parameter values with either no viable FOPT, or for which the variable \protect\py{is_physical} returns \mbox{\protect\py{False}.} We highlight the MAP, MLE and Bayes estimators with a yellow star, a red cross and a purple rhombus, respectively. The dashed orange line is $g_D^\textrm{roll}$,~\Cref{eq:g_roll}.}
    \label{fig:scatter}
\end{figure}

To give a more complete idea of the sampled parameter space, we report in a scatter plot all the sampling points generated during the MCMC in~\Cref{fig:scatter}, with each point coloured according to the log-likelihood value computed by \textsf{PTArcade}. This shows that the employed parametrisation efficiently sampled the phenomenologically relevant region, focusing on the regions with high likelihood but also covering the surrounding areas with lower likelihood values.

\begin{figure}[htb]
    \centering
    \includegraphics[width=0.8\linewidth]{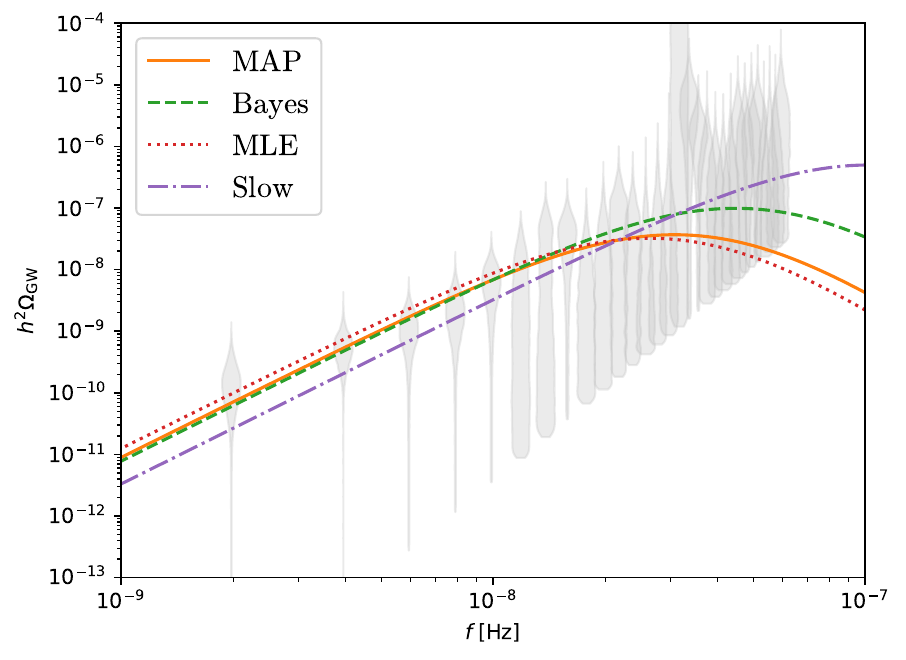}
    \caption{Predicted SGWB spectra for the MAP, MLE and Bayes (central value) estimators in~\Cref{tab:bayes}, compared to the NANOGrav 15-year periodogram. For completeness,  we also show the SGWB of the ``Slow'' point in~\Cref{tab:input} (the signal from the ``Fast'' point is outside of the plotted region).}
    \label{fig:MCMC_GW}
\end{figure}

We plot the predicted SGWB spectra for the MAP, MLE and Bayes estimator parameters, together with the NANOGrav 15-years periodogram in~\Cref{fig:MCMC_GW}; for completeness we also include the SGWB spectrum from the ``Slow'' benchmark point in~\Cref{tab:input}.
A summary of all the discussed points and their thermal parameters as computed by \elena is reported in~\Cref{tab:points}.
\begin{table}[htb]
    \centering
\begin{tabular}{|c|c|c|c|c|c|}
\hline
Quantity & MAP & Bayes & MLE & Slow & Fast \\
\hline
$\lambda_\phi$ & $4.49 \times 10^{-3}$ & $3.21 \times 10^{-3}$ & $1.51 \times 10^{-3}$ & $6 \times 10^{-3}$ & $1.65 \times 10^{-3}$ \\
\hline
$g_D$ & $0.70$ & $0.64$ & $0.53$ & $0.75$ & $0.54$ \\
\hline
$v_\phi^0$ (MeV) & $44.10$ & $1.14 \times 10^{2}$ & $46.72$ & $5 \times 10^{2}$ & $5 \times 10^{2}$ \\
\hline
$m_{\phi}$ (MeV) & $4.18$ & $9.14$ & $2.56$ & $54.77$ & $28.72$ \\
\hline
$m_{Z'}$ (MeV) & $30.75$ & $73.16$ & $24.79$ & $3.75 \times 10^{2}$ & $2.7 \times 10^{2}$ \\
\hline
$T_\textrm{critical}$ (MeV) & $9.50$ & $22.65$ & $7.69$ & $1.16 \times 10^{2}$ & $84.82$ \\
\hline
$T_\textrm{nucleation}$ (MeV) & $0.91$ & $0.90$ & $0.12$ & $13.94$ & $15.27$ \\
\hline
$T_\textrm{percolation}$ (MeV) & $0.74$ & $0.68$ & $0.11$ & $9.61$ & $14.87$ \\
\hline
$T_\textrm{completion}$ (MeV) & $0.68$ & $0.64$ & $0.10$ & $6.39$ & $14.81$ \\
\hline
$T_\textrm{minimal}$ (MeV) & $0.00$ & $0.00$ & $4.13 \times 10^{-2}$ & $0.00$ & $11.55$ \\
\hline
$T_\textrm{reheating}$ (MeV) & $4.46$ & $10.58$ & $4.03$ & $53.84$ & $39.86$ \\
\hline
$P_f^{\rm min}$ & $0.00$ & $2.05 \times 10^{-300}$ & $0.00$ & $6.57 \times 10^{-17}$ & $0.00$ \\
\hline
$\alpha$ & $1.31 \times 10^{3}$ & $5.71 \times 10^{4}$ & $1.98 \times 10^{6}$ & $9.86 \times 10^{2}$ & $50.59$ \\
\hline
$\alpha_\infty$ & $39.79$ & $2.62 \times 10^{2}$ & $1.64 \times 10^{3}$ & $34.45$ & $7.62$ \\
\hline
$\alpha_{\rm eq}$ & $11.07$ & $23.94$ & $47.81$ & $11.79$ & $2.92$ \\
\hline
$\gamma_*$ & $4.26 \times 10^{18}$ & $1.19 \times 10^{18}$ & $6.68 \times 10^{17}$ & $1.69 \times 10^{18}$ & $2.29 \times 10^{16}$ \\
\hline
$\gamma_{\rm eq}$ & $1.15 \times 10^{2}$ & $2.38 \times 10^{3}$ & $4.14 \times 10^{4}$ & $80.67$ & $14.72$ \\
\hline
$R_\textrm{sep}{}_* H_*$ & $0.14$ & $0.23$ & $0.13$ & $0.53$ & $7.13 \times 10^{-3}$ \\
\hline
\end{tabular}
    \caption{Numerical values of the model and FOPT thermal parameters for the specific realisations of the minimal dark sector in~\Cref{eq:lag_UV}: maximum a posterior (MAP), Bayes and maximum likelihood (MLE) estimators from the MCMC sampling discussed in the main text, plus the ``Slow'' and ``Fast'' benchmark points introduced in~\Cref{tab:input}.
    }
    \label{tab:points}
\end{table}

\section{Performance and comparison with \textsf{CosmoTransitions}}
To conclude, we compare \elena with the widely used software \textsf{CosmoTransitions}~\cite{Wainwright:2011kj}, which employs the commonly used bounce formalism (cf.~\Cref{eq:bounce}) to compute the tunnelling action. We compare both the software routines that compute the Euclidean action, as well as the ones that identify the phases of the theory, a step that is required to identify the range of temperatures where a FOPT can happen. All the tests have been performed on a personal machine with an Apple M2 processor.

Concerning the identification of the phases of the theory, we find good agreement among the software, in both the results and the required time, as reported in~\Cref{tab:comparison}.
\begin{table}[htb]
    \centering
\begin{tabular}{|c||c|c||c|c||c|c||c|c||c|c|}
\hline
Point & \multicolumn{2}{c||}{MAP} & \multicolumn{2}{c||}{Bayes} & \multicolumn{2}{c||}{MLE} & \multicolumn{2}{c||}{Slow} & \multicolumn{2}{c|}{Fast} \\
\hline
Software & E & C & E & C & E & C & E & C & E & C \\
\hline
$T_\textrm{crit}$ (MeV) & 9.50  & 9.51 & 22.65 & / & 7.69 & 7.69 & 115.75 & 115.78 & 84.82 & 84.83 \\
\hline
$T_\textrm{min}$ (MeV) & 0 & 0 & 0 & / & 0.04 & 0 & 0 & 0 & 11.55 & 11.48 \\
\hline
time (s) & 0.13 & 0.14 & 0.13 & 0.14 & 0.39 & 0.13 & 0.12 & 0.13 & 0.38 & 0.17 \\
\hline
\end{tabular}
    \caption{Comparison between the values of the critical ($T_\textrm{crit}$) and minimal ($T_\textrm{min}$) temperatures as computed by \elena (E) and \textsf{CosmoTransitions} (C) for the different benchmark points, as well as of the time taken to identify them. For the Bayes estimator, \CT does not identify a second phase; the thermal potential for this point at the critical temperatures identified by \elena is plotted in~\Cref{fig:fast_potential}.}
    \label{tab:comparison}
\end{table}

\begin{figure}[htb]
    \centering
    \includegraphics[width=0.99\linewidth]{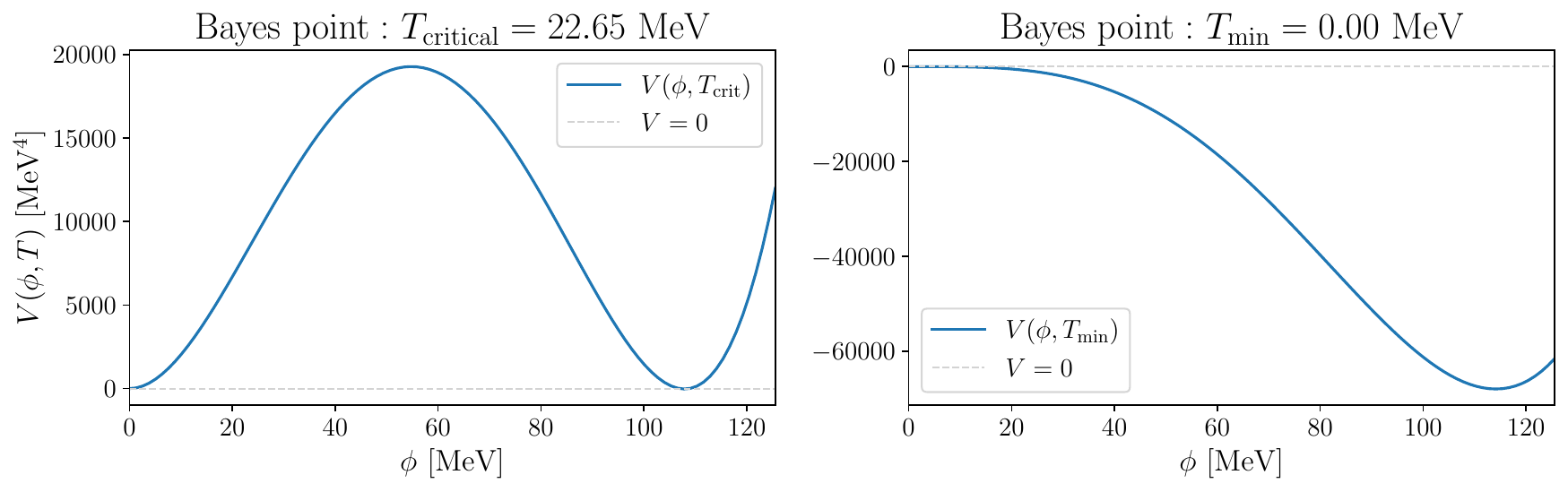}
    \caption{Shape of the thermal potential for the Bayes estimator point at the critical temperature $T_\mathrm{critical}$ (left) and at the minimal one $T_\mathrm{min}$ (right).}
    \label{fig:fast_potential}
\end{figure}

Some differences are observed in the MLE and ``Fast'' points, where \elena takes more time to identify the phases, and quite significantly in the Bayes estimator point, where \textsf{CosmoTransitions} is not able to identify the second phase. We verified graphically that at the critical temperature found by \elena two degenerate local minima exist (cf.~\Cref{fig:fast_potential}) and that the one located away from the origin becomes the absolute minimum at lower temperatures, with a barrier persisting until $T_\textrm{min}$, thus confirming the existence of two phases as correctly identified by \elena.

\begin{figure}[htb]
    \centering
    \includegraphics[width=0.99\linewidth]{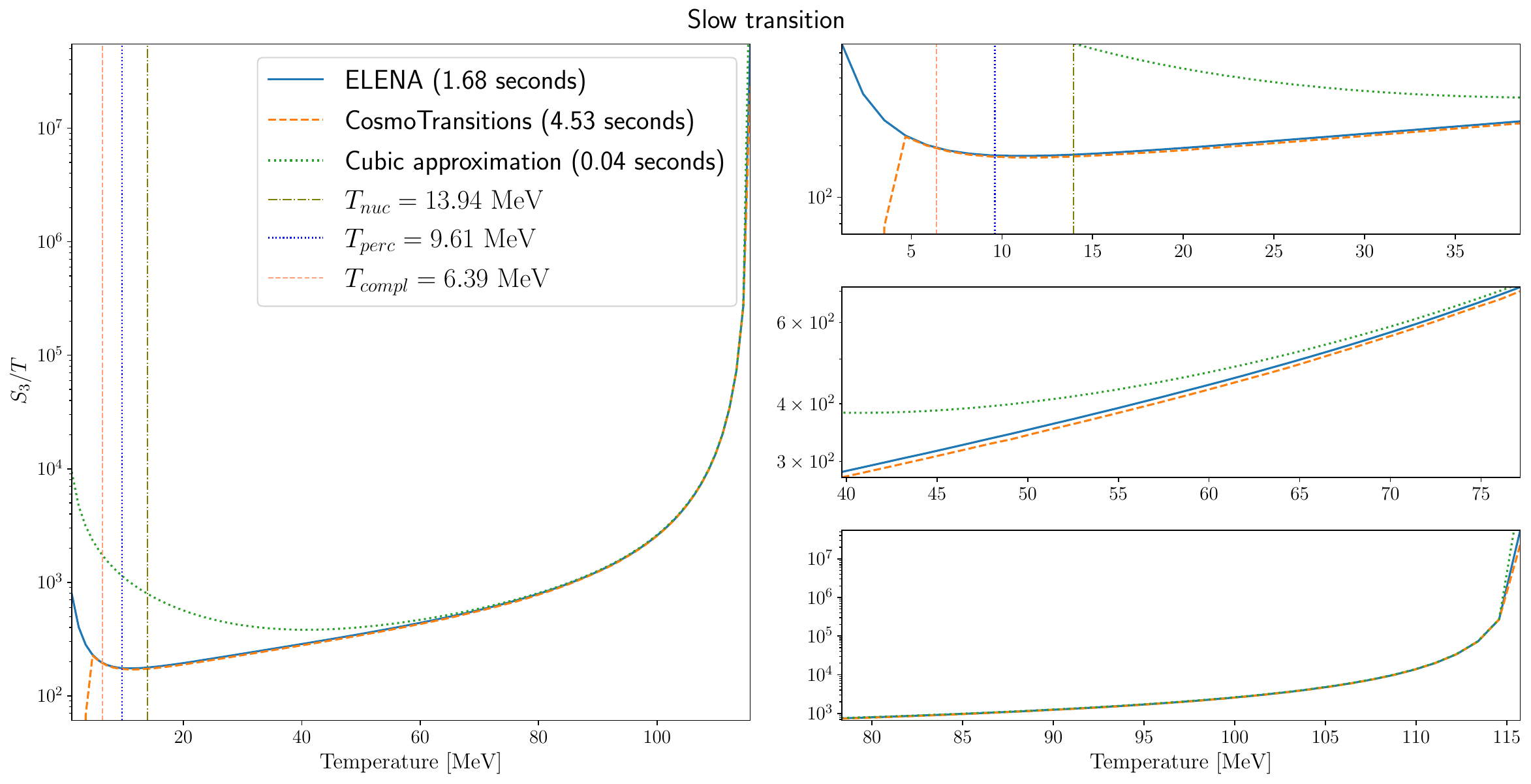}
    \caption{Comparison of the finite-temperature action computation performed by \elena, \CT and by assuming a cubic potential shape (\Cref{eq:cubic}), for the ``Slow'' point in~\Cref{tab:input}. The left panel shows the full range of temperatures between $T_c$ and $T_{min}$, while the right subpanels are zoomed over the low (top), intermediate (middle) and high (bottom) temperature values. The time employed by each algorithm to obtain these results is reported in the legend. For reference, vertical lines represent the nucleation ($T_n$), percolation ($T_p$) and completion ($T_e$) temperatures as dashed, dotted and dot-dashed lines, respectively.}
    \label{fig:CT_action}
\end{figure}

The results for the computation of the action agree as well for most of the parameter space, as demonstrated in~\Cref{fig:CT_action} for the ``Slow'' point in~\Cref{tab:input}. For completeness, we also compare in this figure the results of the numerical computations performed by \elena and \textsf{CosmoTransitions} with the analytic result obtained by fitting the potential to a cubic shape with temperature-dependent polynomial coefficients
\begin{equation}\label{eq:cubic}
    V\left(\phi\right) = m\left(T\right) \phi^2 + \eta\left(T\right) \phi^3 + \lambda\left(T\right) \phi^4.
\end{equation}
This approximation is sometimes employed in the literature because, for a potential described by~\Cref{eq:cubic}, the action can be obtained analytically~\cite{Matteini:2024xvg}. However, it should be noted that~\Cref{eq:cubic} is not guaranteed to provide a good description of the thermal potential in general, as it will be evident in the following.

\begin{figure}[htb]
    \centering
    \includegraphics[width=0.8\linewidth]{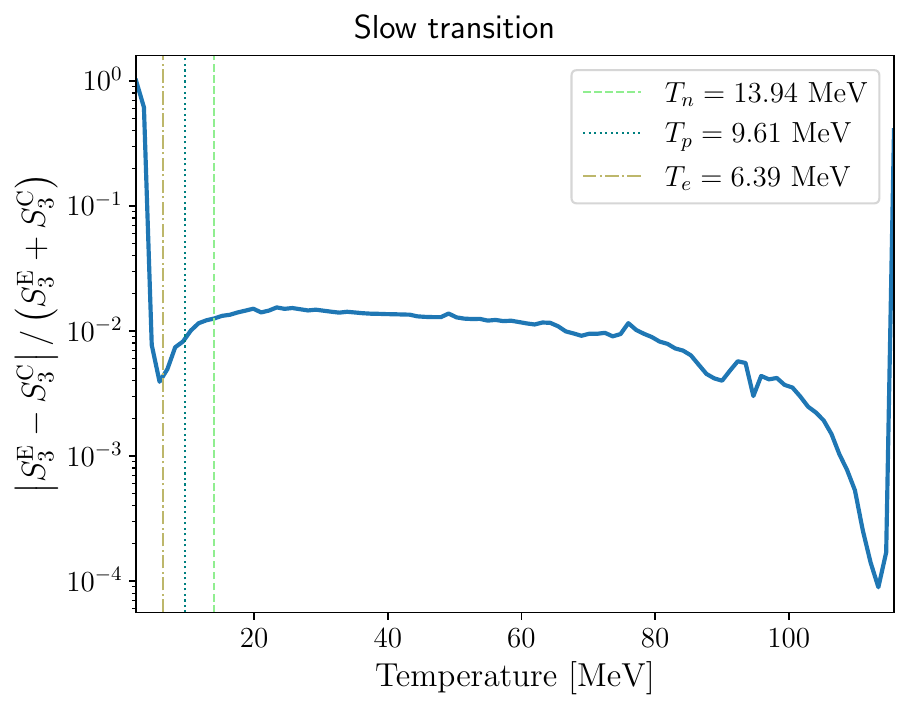}
    \caption{Relative difference of the finite-temperature tunnelling action computed by \elena ($S^\mathrm{E}_3$) and \textsf{CosmoTransitions} ($S^\mathrm{C}_3$), as a function of temperature, for the ``Slow'' point in~\Cref{tab:input}. Vertical lines represent the nucleation ($T_n$), percolation ($T_p$) and completion ($T_e$) temperatures as dashed, dotted and dot-dashed lines, respectively.}
    \label{fig:CT_difference}
\end{figure}

Overall, the two software agree very well, with a relative difference in the results of order $1\%$ or less for most of the range of temperatures, cf.~\Cref{fig:CT_difference}, with the main discrepancies observed at the extreme ends of the temperature range. For $T \sim T_c$ the difference in the action is of order unity, but this has little practical effect since the nucleation rate is infinitesimal at these temperatures. On the other end, for $T \sim T_{min}$, we notice that \textsf{CosmoTransitions} returns negative values of the action, while the results from \elena are always positive, and the shape of the action for very low temperatures does not show any discontinuity, cf.~\Cref{fig:CT_action}.
For completeness, we notice that the estimation of the action obtained with the cubic approximation only works in the high-temperature limit ($T \gtrsim 80$ MeV), where the finite-temperature potential can be correctly described by the polynomial expression in~\Cref{eq:cubic}. For lower temperature values, the conformal shape of the potential becomes relevant, and a simple polynomial expression does not provide a valid fit to it. 

\begin{figure}[htb]
    \centering
    \includegraphics[width=0.8\linewidth]{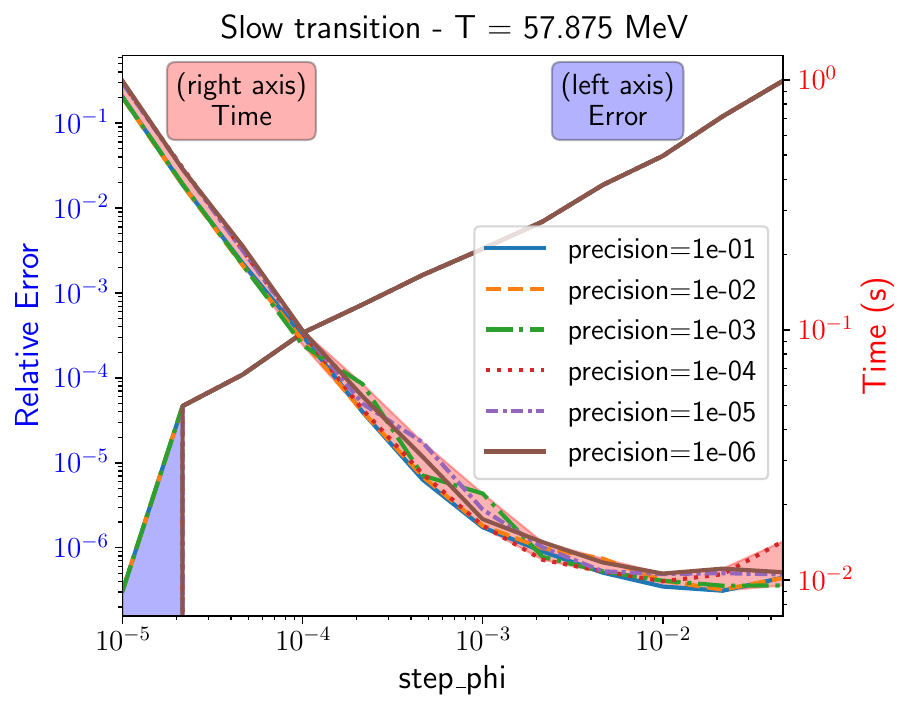}
    \caption{Trade-off between accuracy and computational time in \elena for the computation of the action, using the ``Slow'' point in~\Cref{tab:input} at $T = T_c / 2$ as benchmark scenario. The internal parameters determining the trade-off are \protect\py{step_phi} (varied over the horizontal axis) and \protect\py{precision} (with different values represented by the different line-styles) in the class \protect\py{Vt_vec}. For each combination of parameters, the relative error of the computation is plotted with a light blue background (values to be read on the left vertical axis), while the required computational time is plotted with a light red background (values to be read on the right vertical axis).}
    \label{fig:trade-off}
\end{figure}

To conclude, we comment on the \elena internal parameters that determine the trade-off between numerical accuracy and speed of computation in the evaluation of the action. These are mainly determined by the optional arguments \py{step_phi} and \py{precision} of the class \py{Vt_vec}, as exemplified in the~\Cref{code:action_over_T}. We show the trade-off between error and time in~\Cref{fig:trade-off}, taking as example the ``Slow'' point and evaluating the action at half the critical temperature, $T = 57.875$ MeV. We assume as reference (i.e. most precise) value of the action the one obtained with the smallest choice of both parameters, that is \py{step_phi = 1e-5} and \py{precision = 1e-6}. The change in the running time corresponds to the red shaded region for different precisions, to be read on the right vertical axis, while the relative error corresponds to the blue shaded region, to be read on the left vertical axis.

We notice that the dominant parameter in determining the trade-off is \py{step_phi}, which corresponds to the size of the step in the field dimension, expressed in units of the vev value. This is expected, since this parameter determines how well the potential is sampled over its range, but at the same time, the number of action computations increases with the number of samplings. On the other hand, we find that the relative error does not depend on the \py{precision} parameter for this point, meaning that the first iteration is already enough to converge to the correct $\phi_0$ value; we nevertheless advise against setting too large values for \py{precision}, since different potential shapes may require multiple iterations to converge. On general grounds, we find that the default settings of \py{step_phi = 1e-3} and \py{precision = 1e-3} should provide a valid configuration for most of the users, with a computing time of order $10^{-2}$ seconds and a result that does not differ more than $\mathcal{O}(10^{-3})$ from the most extreme settings. 

\section{Conclusion}
We introduced \elena: \emph{EvaLuator of tunnElliNg Actions}, a Python package designed to study first-order phase transitions generated by scalar potentials in the Early Universe. The core of the package is a vectorised implementation of the tunnelling potential formalism, which is an alternative method to compute the tunnelling action with respect to the widely adopted bounce equation formalism. Contrary to solving the bounce equations of motion, whose solution is a saddle point, in the tunnelling potential approach, the physical action is an absolute minimum in the space of solutions, making the algorithm numerically faster and stable under deviations from the correct solution. To our knowledge, \elena is the first public package that implements this algorithm. 
We tested \elena against the widely used public software \textsf{CosmoTransitions}, finding excellent agreement and faster convergence; in addition, \elena shows strong numerical stability even for very flat potentials, a region where \textsf{CosmoTransitions} can deliver unphysical results due to numerical instabilities of the bounce equation solver.

In addition to providing a novel and efficient algorithm to compute the tunnelling action, we ship \elena with a full suite of additional classes and functions to compute the stochastic gravitational waves background generated in a FOPT. We go beyond current codes by avoiding approximations commonly employed in the literature, for instance, not relying on the bag model, using the percolation temperature instead of the nucleation one, explicitly checking the evolution of the physical volume in false vacuum, and computing the mean bubble separation instead of the linear $\beta$ parameter.

We presented a possible use case for \elena, by interfacing it with the \textsf{PTArcade} software to perform a fit of a minimal dark sector to the NANOGrav 15-year data, being able to sample more than 9 millions points in the underlying MCMC study. We nevertheless stress that the scope of \elena is not limited to Pulsar Timing Array Experiments, but can cover the full range of frequencies explored by current and future experiments testing SGWB from FOPT.

Being a first release, \elena also presents some limitation; most notably, it can only deal with single phase transitions (when only two phases at most are simultaneously present) and only accepts single-field potentials. While we plan to generalise the scope of the software in future releases, we point out to users interested in more complex phase histories that the modularity of \elena (where all individual functions and classes are easily accessible) allows for an easy interface with more specialised software~\cite{Athron:2020sbe, Athron:2024xrh}. We also notice that the tunnelling potential formalism can be readily generalised to multi-field potentials~\cite{Espinosa:2018szu}, and we observe no roadblock into implementing this more general algorithm into our numerical routines, other than the work needed to refactor and extend the code. The extension to multi-field potentials will be the focus of the next development phase.

\section*{Acknowledgements}
The numerical results presented in~\Cref{sec:MCMC} have been obtained by employing the Open Physics Hub HPC facility, hosted at the Department of Physics and Astronomy, University of Bologna. 
ML is funded by the European Union under the Horizon Europe's Marie Sklodowska-Curie project 101068791 — NuBridge. F. C. acknowledges support from the FORTE
project CZ.02.01.01/00/22 008/0004632 co-funded by
the EU and the Ministry of Education, Youth and
Sports of the Czech Republic. JHZ is supported by the National Science Centre, Poland (research grant No. 2021/42/E/ST2/00031). In the initial stages of this project, the research of FC, JHZ, SP and SRA received support from the European Union’s
Horizon 2020 research and innovation programme under
the Marie Sklodowska-Curie grant agreement No 860881-HIDDeN. ML thanks Fermilab for hosting him during the development of part of this work.

\appendix

\section{Installation example using the \textsf{conda} package manager}
\elena is designed to be as much a self-contained Python package as possible, only requiring a small number of dependencies. Specifically, it depends on \textsf{CosmoTransitions} for constructing the finite-temperature class model, plus the \textsf{NumPy} and \textsf{SciPy} libraries to perform numerical operations. To fully run the notebook examples shipped with the software some additional dependencies are required, specifically for data visualisation and interactive sessions, but we stress that these are not required to run \elena itself.

In the following, we show how to use the \textsf{conda} package manager to create a self-contained environment where \elena can be executed. The user simply needs to run the following commands in the terminal:
\begin{lstpy}
conda create --name elena -c conda-forge python numpy scipy cosmoTransitions matplotlib ipykernel natpy la_forge
conda activate elena
\end{lstpy}
At this point, a \textsf{conda} environment containing the required dependencies will be created and activated. The next step is to download the \elena code:
\begin{lstpy}
git clone https://github.com/michelelucente/ELENA
cd ELENA 
\end{lstpy}
\elena is now present on the local machine, and the user can test it by running for instance the \textsf{Jupyter} notebook located in \py{'./examples/phase_transition.ipynb'}.

\bibliographystyle{JHEP}
\bibliography{bibliography.bib}

\end{document}